\documentclass[a4paper,11pt]{article}

\pdfoutput=1

\usepackage{amssymb,amsmath,amsfonts,jheppub}
\usepackage{mathrsfs,mathtools}
\usepackage{ccaption} 
\usepackage{graphicx}
\usepackage{tikz}
\usetikzlibrary{decorations.pathmorphing,calc}

\usepackage[all]{hypcap}

\def\clock{{\count0=\time
           \divide\count0 60
           \ifnum\count0<10 0\fi\the\count0
           \multiply\count0 -60 \advance\count0 \time
           :\ifnum\count0<10 0\fi \the\count0
         }}
\newcommand{\timestamp}{{\small\vbox{\hbox{\tt\jobname.tex}
\hbox{\the\day/\the\month/\the\year, \clock}}}}

\usepackage{amsmath,amssymb}
\usepackage{verbatim}
\usepackage{graphicx}
\usepackage{hyperref}
\usepackage{color}

\DeclareFontFamily{OT1}{rsfs}{}
\DeclareFontShape{OT1}{rsfs}{m}{n}{ <-7> rsfs5 <7-10> rsfs7 <10->rsfs10}{} 
\DeclareMathAlphabet{\mycal}{OT1}{rsfs}{m}{n}

\newcommand{\unity}{1\hspace{-0.243em}\text{l}}

\newcommand{\be}[1]{ \begin{equation}\label{#1} }
\newcommand{\ee}{\end{equation}}
\newcommand{\bea}[1]{\begin{eqnarray}\label{#1} }
\newcommand{\eea}{\end{eqnarray}}

\newcommand{\tr}{\textrm{tr}}

\newcommand{\eq}[2]{\begin{equation} #1 \label{#2} \end{equation}}
\newcommand{\eps}{\varepsilon}

\newcommand{\de}{\delta}

\newcommand{\la}{\lambda}

\newcommand{\De}{\Delta}

\DeclareMathOperator{\extdm}{d}
\newcommand{\extd}{\extdm \!}

\subheader{TUW--14--15}
\title{Flat space (higher spin) gravity with chemical potentials}
\author[a]{Michael Gary,}
\author[a]{Daniel Grumiller,}
\author[a,b]{Max Riegler,}
\author[a]{and Jan Rosseel}
\affiliation[a]{Institute for Theoretical Physics, Vienna University of Technology\\
Wiedner Hauptstrasse 8-10/136, A-1040 Vienna, Austria}
\affiliation[b]{Yukawa Institute for Theoretical Physics (YITP), Kyoto University, Kyoto 606-8502, Japan}
\emailAdd{mgary@hep.itp.tuwien.ac.at}
\emailAdd{grumil@hep.itp.tuwien.ac.at}
\emailAdd{rosseelj@hep.itp.tuwien.ac.at}
\emailAdd{rieglerm@hep.itp.tuwien.ac.at}
\abstract{We introduce flat space spin-3 gravity in the presence of chemical potentials and discuss some applications to flat space cosmology solutions, their entropy, free energy and flat space orbifold singularity resolution. Our results include flat space Einstein gravity with chemical potentials as special case.
We discover novel types of phase transitions between flat space cosmologies with spin-3 hair and show that the branch that continuously connects to spin-2 gravity becomes thermodynamically unstable for sufficiently large temperature or spin-3 chemical potential.}
\arxivnumber{1411.3728}

\begin{document}

\maketitle

\section{Introduction}\label{se:1}

Chemical potentials are ubiquitous in physics (and chemistry) ever since their introduction by Gibbs. In gauge theories chemical potentials $\mu$ are usually introduced by giving the $0$-component of the gauge connection a vacuum expectation value (see e.g.~\cite{Kapusta}). 
\eq{
A_0 \to A_0 + \mu
}{eq:intro1}

In the present work we are interested in flat space higher spin gravity with chemical potentials. We start by summarizing briefly selected recent developments that provide the motivation for our work. This is not meant to be a comprehensive review, but merely serves to put our motivation in the context of the current research directions.

Higher spin gravity in Anti-de~Sitter (AdS) has led to numerous holographic studies, many of which were inspired by the seminal work by Klebanov and Polyakov \cite{Klebanov:2002ja,Mikhailov:2002bp,Sezgin:2002rt} who conjectured a holographic correspondence between the $O(N)$ vector model in three dimensions and Fradkin--Vasiliev higher spin gravity on AdS$_4$ \cite{Fradkin:1987ks,Fradkin:1986qy,Vasiliev:1990en} (see \cite{Sagnotti:2010at,Vasiliev:2012vf,Didenko:2014dwa} for reviews and \cite{Giombi:2009wh,Giombi:2010vg,Koch:2010cy,Giombi:2011ya,Douglas:2010rc} for some key developments). One of the attractive features of higher spin holography for the purpose of checking the holographic principle is that it is a weak/weak correspondence, i.e., relates higher spin gravity theories to very simple conformal field theories (CFTs) \cite{Maldacena:2011jn,Maldacena:2012sf}. By contrast, the usual AdS/CFT correspondence \cite{Maldacena:1997re,Gubser:1998bc,Witten:1998qj} is a weak/strong correspondence, which makes 
it useful for applications, but harder to check in detail, since calculations are often feasible only on one side of the correspondence.

We pause now briefly our mini-history to reconsider our goal of introducing chemical potentials in flat space higher spin gravity. It may not be immediately clear how to do this technically. However, when the theory allows a (classically) equivalent reformulation as gauge theory one can again use a prescription like \eqref{eq:intro1}.

Gravity, including higher spin gravity, in three dimensions does allow for such a reformulation, namely as Chern--Simons theory \cite{Achucarro:1986vz,Witten:1988hc,Blencowe:1988gj}. Indeed, exploiting this formulation chemical potentials were introduced in spin-3 AdS gravity in the past few years, first in the form of new black hole solutions with spin-3 fields by Gutperle and Kraus \cite{Gutperle:2011kf} (see also \cite{Ammon:2011nk}), next perturbatively in the spin-3 chemical potential \cite{Compere:2013gja}, then to all orders by Comp\`ere, Jottar and Song \cite{Compere:2013nba} and independently by Henneaux, Perez, Tempo and Troncoso \cite{Henneaux:2013dra}. A comprehensive recent discussion of higher spin black holes with chemical potentials is provided in \cite{Bunster:2014mua}. The discussion so far was focused mostly on AdS and holographic aspects thereof \cite{Gaberdiel:2010pz}, see \cite{Ammon:2012wc,Gaberdiel:2012uj,Perez:2014pya,Afshar:2014rwa} for reviews.

As advocated in \cite{Gary:2012ms}, higher spin gravity has turned out to be a fertile ground for non-AdS holography, without the necessity for additional exotic matter degrees of freedom, including Lobachevsky holography \cite{Afshar:2012nk,Afshar:2012hc}, Lifshitz holography \cite{Gutperle:2013oxa,Gary:2014mca}, de~Sitter holography \cite{Krishnan:2013zya} and flat space holography \cite{Afshar:2013vka,Gonzalez:2013oaa}. This is not only of interest in its own right, but particularly for verifying the generality of the holographic principle \cite{'tHooft:1993gx,Susskind:1995vu}, which should apply beyond AdS/CFT if it is a true aspect of Nature. 

Three-dimensional flat space higher spin gravity is especially remarkable, since in higher dimensions massless interacting higher spin theories in flat space are forbidden by various no-go results \cite{Coleman:1967ad,Aragone:1979hx,Weinberg:1980kq} (see \cite{Bekaert:2010hw} for a nice summary). It is therefore interesting that in three dimensions such theories exist \cite{Afshar:2013vka,Gonzalez:2013oaa}, though unitarity again poses strong constraints on the theory and rules out the simplest realizations of unitary flat space higher spin gravity \cite{Grumiller:2014lna}.

The numerous recent advances in (three-dimensional) higher spin gravity are matched by exciting developments in (three-dimensional) flat space holography. Starting from the Barnich--Comp\`ere boundary conditions \cite{Barnich:2006av}, some key developments were the BMS/CFT or BMS/GCA correspondences\footnote{%
BMS stands for Bondi--van~der~Burg--Metzner--Sachs \cite{Bondi:1962,Sachs:1962}, the asymptotic symmetry algebra of flat spacetimes at null infinity, and GCA for Galilean conformal algebra \cite{Bagchi:2009my}. 
} \cite{Barnich:2010eb,Bagchi:2010zz}, the flat space chiral gravity proposal \cite{Bagchi:2012yk}, the counting of flat space cosmology microstates \cite{Bagchi:2012xr,Barnich:2012xq}, the existence of phase transitions between flat space cosmologies and hot flat space \cite{Bagchi:2013lma} and numerous other aspects of flat space holography \cite{Barnich:2012aw,Bagchi:2012cy,Bagchi:2013bga,
Costa:2013vza,Fareghbal:2013ifa,Krishnan:2013tza,Krishnan:2013wta,Bagchi:2013qva,Detournay:2014fva,Barnich:2014kra,Barnich:2014cwa,Riegler:2014bia,Fareghbal:2014qga}.

In AdS$_3$/CFT$_2$ it is rewarding to study Ba\~nados--Teitelboim--Zanelli (BTZ) black holes \cite{Banados:1992wn,Banados:1992gq}. The flat space analogue of these objects are flat space cosmologies \cite{Cornalba:2002fi,Cornalba:2003kd}. Much like it is possible to consider BTZ black holes or their higher spin versions with chemical potentials switched on, it is plausible that there should be a flat space counterpart thereof, both in flat space gravity and in flat space higher spin gravity. 

Flat space higher spin gravity in three dimensions combines all these research avenues and may serve to gain a better and deeper understanding of higher spin gravity, flat space holography, microscopic aspects of flat space cosmologies, string theory in the tensionless limit and, more broadly, quantum gravity and the holographic principle itself. 

In the present work we consider specifically spin-3 gravity in flat space. Our main goal is to introduce chemical potentials for the spin-2 and spin-3 field in flat space, and to address some of their consequences, in particular the entropy and free energy of flat space cosmologies with spin-3 charges.
Technically, we do this by working in the Chern--Simons formulation of spin-3 gravity and introducing chemical potentials as in \eqref{eq:intro1}, i.e., by deforming the zero-component of the gauge connection, analog to \cite{Bunster:2014mua}.

One of the most surprising results that we find is that the ``physical'' branch that connects continuously to spin-2 physics becomes thermodynamically unstable at large temperature or large spin-3 chemical potential. The phase transition can be of first or zeroth order, which differs qualitatively from the situation in AdS, where the corresponding phase transitions discovered so far were of zeroth order \cite{David:2012iu,Chen:2012ba,Ferlaino:2013vga}.

This paper is organized as follows. 
In section \ref{se:2} we review aspects of flat space spin-2 and spin-3 gravity.
In section \ref{se:3} we include chemical potentials and present the main results. 
In section \ref{sec:E} we display our results applied to flat space Einstein gravity with chemical potentials.
In section \ref{se:4} we discuss some applications to flat space cosmologies, calculate their entropy and free energy, discover novel types of phase transitions, remark on flat space orbifold singularity resolution in spin-3 gravity and mention further developments and open issues.

\section{Flat space higher spin gravity}\label{se:2}

In this section we review results of flat space higher spin gravity and present them in a way that is considerably simpler than in the original publications. In section \ref{se:2.1} we recall the Chern--Simons formulation in terms of an isl$(3)$ connection. In section \ref{se:2.2} we display a recent representation of this connection reminiscent of similar representations in the AdS case. In section \ref{se:2.3} we present the canonical charges and their algebra. In section \ref{se:2.4} we provide simple formulas for metric and spin-3 field by means of a twisted trace.

\subsection{Chern--Simons formulation}\label{se:2.1}

Like in the AdS case, it is very convenient to use the Chern--Simons formulation of the theory. The Chern--Simons action
\eq{
I[A]=\frac{k}{4\pi}\int\,\langle\textrm{CS}(A)\rangle
}{eq:cp1}
contains a coupling constant $k$ (the Chern--Simons level; $k=1/(4G_N)$, where $G_N$ is Newton's constant) and the Chern--Simons 3-form
\eq{
\textrm{CS}(A) = A\wedge \extd A + \tfrac23\,A\wedge A \wedge A\,.
}{eq:cp2}
The bilinear form $\langle\cdot\,,\,\cdot\rangle$ will be specified below.
In the present work the connection will always be isl$(3)$, i.e., it can be decomposed into a linear combination of isl$(3)$ generators $G_n$ as
\eq{
A=\sum_{n=1}^{16} A^n G_n=\sum_{n=-1}^1 \big(A^n_L L_n + A^n_M M_n \big) + \sum_{n=-2}^2 \big(A^n_U U_n + A^n_V V_n\big)
}{eq:cp3}
with the generators $L_n, M_n, U_n, V_n$ obeying the isl$(3)$ algebra.
\begin{subequations}
\label{eq:FSHSG14}
\begin{align}
[L_n,\, L_m]&=(n-m)L_{n+m}\\
[L_n,\, M_m]&=(n-m)M_{n+m}\\
[L_n,\, U_m]&=(2n-m)U_{n+m}\\
[L_n,\, V_m]&=(2n-m)V_{n+m}=[ M_n,\, U_m] \\
[U_n,\, U_m]&=\sigma\,(n-m)(2n^2+2m^2-nm-8)L_{n+m}\\
[U_n,\, V_m]&=\sigma\,(n-m)(2n^2+2m^2-nm-8)M_{n+m}
\end{align}
\end{subequations}
The $L_n$ generate (Lorentz-)rotations, $M_n$ generate translations, and $U_n$, $V_n$ generate associated spin-3 transformations. The factor $\sigma$ fixes the overall normalization of the spin-3 generators $U_n$ and $V_n$. In the present work we choose\footnote{%
The minus sign in \eqref{eq:sigma} guarantees that the generators $L_n$ and $U_n$ form an sl$(3,\mathbb{R})$ subalgebra, corresponding to the maximally non-compact real form of $A_2$.
}
\eq{
\sigma=-\frac13\,.
}{eq:sigma}

It is also noteworthy that one can equip the algebra \eqref{eq:FSHSG14} naturally with a ${\mathbb Z}_2$ grading so that the generators $L_n$, $U_n$ are even and $M_n$, $V_n$ are odd. Then even with even gives even, even with odd gives odd and odd with odd vanishes. If one constructs isl$(3)$ as an \.In\"on\"u--Wigner contraction from so$(2,2)$ it is possible to do so introducing a Grassmann parameter $\epsilon$, and the contraction then consists in dropping all expressions quadratic in $\epsilon$; the $\epsilon$-independent generators are then the even generators and the generators linear in $\epsilon$ are the odd generators \cite{Krishnan:2013wta}. 

We exploit this Grassmann trick in appendix \ref{app:A.1} to define the generators. Moreover, we use it to define a (twisted) trace over a product of $k$ isl$(3)$ generators $G_{n_i}$ (with $i=1\dots k$) that is useful to construct the spin-2 and spin-3 fields from the isl$(3)$ connection.
\eq{
\widetilde\tr\Bigg(\prod_{i=1}^k G_{n_i}\Bigg) := \frac12\,\tr\Bigg(\prod_{i=1}^k\bigg(\frac{\extd}{\extd\epsilon}\,G_{n_i}\times \gamma^\ast\bigg)\Bigg)
}{eq:app13}
The right hand side contains the usual matrix trace and involves the matrix \eqref{eq:app6}.
Here are some relevant properties of the twisted trace:
\begin{itemize}
 \item {\bf Oddness.} The twisted trace vanishes identically if at least one of the generators $G_{n_i}$ is even, i.e., one of the generators $L_n$ or $U_n$. Therefore, the only non-vanishing twisted traces involve exclusively the odd generators $M_n$ and $V_n$.
 \item {\bf Relation to matrix trace.} If $k$ is even, then all factors of $\gamma^\ast$ cancel and the twisted trace is equivalent to the ordinary trace, upon taking into account the oddness property and up to a factor $\tfrac12$. If $k$ is odd, then one factor of $\gamma^\ast$ remains, which ensures that the twisted trace does not vanish identically for all odd numbers of odd generators (the ordinary trace, however, does vanish for all odd numbers of odd generators, essentially due to the vanishing trace of the Pauli matrix $\sigma_3$).
 \item {\bf Relation to sl${\boldsymbol (3)}$ trace.} If we consider just the sl$(3)$ block of the generators then the twisted trace is equivalently defined as the matrix trace over the products of the corresponding sl$(3)$ blocks, again upon taking into account the oddness property. For the sake of this property we introduced the factor $\tfrac12$ in the definition \eqref{eq:app13}.
\end{itemize}
We shall employ the twisted trace \eqref{eq:app13} to define the spin-2 and spin-3 fields.

With respect to the above generators the (degenerate) bilinear form is given by
\eq{
\langle L_m,\, M_n\rangle=-2\eta_{mn}\qquad\langle U_m,\, V_n
\rangle=\tfrac23\,\mathcal{K}_{mn}\,.
}{eq:angelinajolie}
Here $\eta_{mn}$ given by $\eta = \textrm{antidiag}\, (1, \,-\tfrac12, \,1)$ is proportional to the sl$(2)$ Killing form and the sl$(3)$ part is given by $\mathcal K = \textrm{antidiag}\, (12, \,-3, \,2, \,-3, \,12)$, both of which have non-zero entries only on the anti-diagonal. The bilinear form can be represented as a trace as follows [again using the matrix \eqref{eq:app6}]:
\eq{
\langle G_{n_1} G_{n_2} \rangle = \widehat\tr\big(G_{n_1} G_{n_2}\big) := \frac{\extd}{\extd \epsilon}\,\tfrac14\,\tr\big(G_{n_1} G_{n_2}\gamma^\ast\big)\big|_{\epsilon=0}
}{eq:cp64}
We shall refer to this trace as ``hatted trace'' to discriminate it from the twisted trace \eqref{eq:app13} and the ordinary matrix trace.

\subsection{Spin-3 flat space connection}\label{se:2.2}

Explicit expressions for isl$(3)$ connections that obey asymptotically flat boundary conditions were established independently in \cite{Afshar:2013vka} and \cite{Gonzalez:2013oaa} (see also \cite{Afshar:2013bla}). However, we shall not use either of these expressions, but use instead the one introduced in \cite{Barnich:2014cwa} since it is considerably simpler. Namely, we represent the connection $A$ as gauge transformation of another connection $a$ with very simple properties and simple gauge group element $b$:
\eq{
A = b^{-1}\,\big(\extd + a\big)\,b
}{eq:cp4}
with
\eq{
b = \exp{\big(\tfrac{r}{2}\, M_{-1}\big)}
}{eq:cp5}
and
\eq{
a = a_u(u,\varphi)\,\extd u + a_\varphi(u,\varphi)\,\extd\varphi
}{eq:cp6}
The form \eqref{eq:cp4} is reminiscent of the similarly useful form of the AdS connection in spin-2 gravity (see e.g.~\cite{Banados:1998gg,Carlip:2005zn}) and higher spin gravity \cite{Henneaux:2010xg,Campoleoni:2010zq}.

We use coordinates $u$, $r$ and $\varphi\sim\varphi+2\pi$ adapted to flat space in (outgoing) Eddington--Finkelstein coordinates. As we shall see below, the background line-element in the absence of chemical potentials is then given by
\eq{
\extd s^2 = -\extd u^2-2\extd r\extd u+r^2\,\extd\varphi^2\,.
}{eq:cp11}
The manifold is topologically a filled cylinder. The asymptotic boundary cylinder (corresponding to null infinity) is reached in the limit where the radial coordinate $r$ tends to infinity.

The boundary conditions of \cite{Afshar:2013vka,Gonzalez:2013oaa} simplify to conditions on $a_u(u,\varphi)$ and $a_\varphi(u,\varphi)$.
\begin{subequations}
 \label{eq:cp7}
\begin{align}
 a_u &= M_+ -\frac{{\cal M}}{4}\,M_- + \frac{{\cal V}}{2}\,V_{-2}\\
 a_\varphi &= L_+ -\frac{{\cal M}}{4}\,L_- + \frac{{\cal V}}{2}\,U_{-2} - \frac{{\cal N}}{2}\, M_- + {\cal Z}\,V_{-2}
\end{align}
\end{subequations}
Note that any connection $A$ with the properties \eqref{eq:cp4}-\eqref{eq:cp7} automatically solves the Chern--Simons field equations
\eq{
F = \extd A + [A,\,A] = 0
}{eq:cp8}
provided the state-dependent functions in \eqref{eq:cp7} are constrained as follows.
\eq{
\dot {\cal M} = \dot {\cal V} = 0\qquad \dot{\cal N} = \tfrac12\,{\cal M}^\prime \qquad  \dot{\cal Z} = \tfrac12\,{\cal V}^\prime
}{eq:cp9}
Dots (primes) denote derivatives with respect to retarded time $u$ (angular coordinate $\varphi$). 
The constraints \eqref{eq:cp9} are solved in terms of four arbitrary functions of the angular coordinate $\varphi$, all of which appear in the canonical charges \cite{Afshar:2013vka}.
\eq{
{\cal M} = {\cal M}(\varphi)\qquad{\cal V} = {\cal V}(\varphi)\qquad{\cal N} = {\cal L}(\varphi)+\tfrac{u}{2}\,{\cal M}^\prime(\varphi) \qquad{\cal Z} = {\cal U}(\varphi)+\tfrac{u}{2}\,{\cal V}^\prime(\varphi)
}{eq:cp10}

\subsection{Canonical charges and their algebra}\label{se:2.3}

The canonical charges of Brown--Henneaux type \cite{Brown:1986nw} are constructed in the usual way \cite{Henneaux:2010xg,Campoleoni:2010zq}. We use here the notation of \cite{Afshar:2012nk}.

The first step is to identify all gauge transformations that preserve the flat space boundary conditions. This was done already in \cite{Afshar:2013vka,Gonzalez:2013oaa}. We rephrase now these results in terms of the new representation of the connection \eqref{eq:cp4}-\eqref{eq:cp6}. To this end, we similarly define the gauge parameter as
\eq{
\varepsilon=b^{-1}\, \varepsilon^{(0)}\, b
}{eq:cp23}
with the same group element \eqref{eq:cp5} as before. The results of \cite{Afshar:2013vka,Gonzalez:2013oaa} for the boundary condition preserving gauge transformations then translate into the following expression:\footnote{%
There are three differences to the results in \cite{Afshar:2013vka}, whose conventions we use: 1.~due to the convenient representation \eqref{eq:cp23} with \eqref{eq:cp5} we do not have any $r$-dependent terms, which are automatically generated through the Baker--Campbell--Hausdorff formula, 2.~we have corrected three numerical coefficients, which all differ by factor of $-3$ from the expressions given in \cite{Afshar:2013vka}, namely the coefficients of the ${\cal V}$- and ${\cal Z}$-terms in the components $L_-$ and $M_-$, and 3.~we have rescaled $\tau$ by a factor of $2$ to make the results look more symmetric.
We note finally that we use $\epsilon$ in two ways in this paper, as Grassmann parameter and as function in the boundary condition preserving gauge transformations \eqref{eq:cp24}, but we believe that the meaning should always be clear from the context.
}
\begin{align}
 \varepsilon^{(0)} &= \epsilon\,L_+ - \epsilon^\prime\,L_0 + \tfrac12\,\big(\epsilon''- \tfrac12{\cal M}\epsilon-8{\cal V}\chi\big)\,L_- \nonumber\\
& \quad + \tau\,M_+ - \tau^\prime\, M_0 + \tfrac12\,\big(\tau''-\tfrac12{\cal M}\tau-{\cal N}\epsilon-8{\cal V}\kappa-16{\cal Z}\chi\big)\,M_- \nonumber \\
&\quad + \chi\,U_2 - \chi^\prime \,U_1 + \tfrac12\,\big(\chi''-{\cal M}\chi\big)\,U_0 - \tfrac16\,\big(\chi'''-\tfrac52{\cal M}\chi^\prime-{\cal M}^\prime\chi\big) \,U_{-1} \nonumber \\
&\qquad + \tfrac{1}{24}\,\big(\chi''''-4{\cal M}\chi''-\tfrac72{\cal M}^\prime\chi^\prime-{\cal M}''\chi+\tfrac32\,{\cal M}^2\chi+12{\cal V}\epsilon\big)\,U_{-2} \nonumber \\
& \quad + \kappa\,V_2 - \kappa^\prime \,V_1 + \tfrac12\,\big(\kappa''-{\cal M}\kappa-2{\cal N}\chi\big)\,V_0 \nonumber \\
& \qquad - \tfrac16\, \big(\kappa'''-\tfrac52{\cal M}\kappa^\prime-{\cal M}^\prime\kappa-5{\cal N}\chi^\prime-2{\cal N}^\prime\chi\big)\,V_{-1}  + \tfrac{1}{24}\,\big(\kappa''''-4{\cal M}\kappa''-\tfrac72{\cal M}^\prime\kappa^\prime \nonumber \\
& \qquad -{\cal M}''\kappa+\tfrac32{\cal M}^2\kappa - 8{\cal N}\chi''-7{\cal N}^\prime\chi^\prime-2{\cal N}''\chi+6{\cal MN}\chi + 12{\cal V}\tau + 24{\cal Z}\epsilon\big)\,V_{-2}
\label{eq:cp24}
\end{align}
The functions $\epsilon$, $\sigma$, $\chi$ and $\rho$ depend on $\varphi$ only, and we have the relations $\tau=\sigma+u\epsilon^\prime$ and $\kappa=\rho+u\chi^\prime$.
When acting with such a gauge transformation on an isl$(3)$ connection $A$ with the properties \eqref{eq:cp4}-\eqref{eq:cp10} 
\eq{
\delta_\varepsilon A = \extd\varepsilon + [A,\varepsilon]
}{eq:cp27}
the gauge transformed connection $\widehat A = A + \delta_\varepsilon A$ also has the properties \eqref{eq:cp4}-\eqref{eq:cp10}, in general with some shifted values for the state dependent functions, $\widehat{\cal M}={\cal M}+\delta_{\varepsilon}{\cal M}$, and similarly for ${\cal N}$, ${\cal V}$ and ${\cal Z}$.

The canonical charges also follow the general prescription of (non-)AdS holography summarized in \cite{Afshar:2012nk}. Their field variation, also known as the canonical currents, is given by
\eq{
\delta Q[\varepsilon] = \frac{k}{2\pi}\,\oint\extd\varphi\,\widehat\tr\big(\varepsilon^{(0)}\delta a_\varphi\big)
}{eq:cp22}
Note the appearance of the hatted trace \eqref{eq:cp64}.

Inserting the expressions \eqref{eq:cp7} and \eqref{eq:cp24} into the canonical currents \eqref{eq:cp22} yields
\eq{
\delta Q[\varepsilon] = \frac{k}{2\pi}\,\oint\extd\varphi\,\big(\epsilon\,\delta{\cal L}+\tfrac12\sigma\,\delta{\cal M}+8\chi\,\delta{\cal U}+4\rho\,\delta{\cal V}\big)
}{eq:cp25}
It is now evident that the canonical currents can be integrated in field space to canonical boundary charges
\eq{
 Q[\epsilon,\tau,\chi,\kappa] = \frac{k}{2\pi}\,\oint\extd\varphi\,\big(\epsilon\,{\cal L}+\tfrac12 \sigma\,{\cal M}+8\chi\,{\cal U}+4\rho\,{\cal V}\big)\,.
}{eq:cp26}
The canonical charges are integrable, finite and conserved in (retarded) time, $\partial_u Q = 0$.

The algebra of the canonical charges was derived classically \cite{Afshar:2013vka,Gonzalez:2013oaa} and quantum-mechanically \cite{Afshar:2013vka} starting from the Poisson-bracket algebra of the canonical charges \eqref{eq:cp26} and then expanding in Fourier modes, e.g.
\eq{
\mathcal{L}(\varphi) \propto \sum_{n \in \mathbb{Z}} \mathcal{L}_n e^{-i n \varphi}
}{eq:FSHSG7}
and similarly for the other three state-dependent functions appearing in \eqref{eq:cp26}. After a suitable shift of the zero mode ${\cal M}_0\to{\cal M}_0+\tfrac k2$ and converting Poisson-brackets into commutators one obtains finally the asymptotic symmetry algebra as a commutator algebra of the modes ${\cal L}_n$, ${\cal M}_n$, ${\cal U}_n$ and ${\cal V}_n$.
It is an \.In\"on\"u--Wigner contraction of two copies of the ${\cal W}_3$ algebra, with the following non-vanishing commutators.
 \begin{subequations}
 \label{eq:FSHSG10}
\begin{align}
[L_n,\, L_m] &= (n-m) L_{n+m} \\
[L_n,\, M_m] &= (n-m) M_{n+m} + k\, (n^3 - n) \delta_{n+m,\,0}  \\
[L_n,\, U_m] &= (2n-m) U_{n+m}  \\
[L_n,\, V_m] &= (2n-m) V_{n+m}  \\
[M_n,\, U_m] &= (2n-m) V_{n+m}  \\
[U_n,\, U_m] &= -\tfrac13\,(n-m)(2 n^2 + 2 m^2 - nm -8) L_{n+m} \nonumber \\ & \quad
-\frac{16}{3k}  (n-m)\Lambda_{n+m}  + \frac{88}{45 k^2}  (n-m)\Theta_{n+m} \label{eq:FSHSG20} \\
[U_n,\, V_m] &= -\tfrac13\,(n-m)(2 n^2 + 2 m^2 - nm -8) M_{n+m} \nonumber \\ &\quad
- \frac{8}{3k}  (n-m) \Theta_{n+m} -\frac k3\, n(n^2-1)(n^2-4) \delta_{n+m,\,0} 
\end{align}
We used the definitions of bi-linears in the generators
\eq{
\Theta_m=\sum_p M_p\, M_{m-p} \qquad \Lambda_m=\sum_p\colon\!L_p\, M_{m-p}\colon -\tfrac{3}{10}(m+2)(m+3)M_m
}{eq:FSHSG8}
\end{subequations}
where normal ordering is defined by
$\colon\!L_n\, M_m\colon = L_n\, M_m$ if $n<-1$ and
$\colon\!L_n\, M_m\colon = M_m\, L_n$ otherwise. It is interesting to note that the algebra \eqref{eq:FSHSG10}, with some standard assumptions, does not have unitary highest weight representations for non-vanishing $k$ \cite{Grumiller:2014lna}.

\subsection{Metric and spin-3 field}\label{se:2.4}

The metric in AdS higher spin gravity is usually defined as trace over the zuvielbein \cite{Henneaux:2010xg,Campoleoni:2010zq}. In flat space higher spin gravity the line-element takes the form
\eq{
\extd s^2 = g_{\mu\nu}\,\extd x^\mu\extd x^\nu = \eta_{mn} A^m_M A^n_M  + \mathcal{K}_{mn} A^m_V A^n_V 
}{eq:cp12}
which for the connection \eqref{eq:cp4}-\eqref{eq:cp10} simplifies to
\eq{
g_{\mu\nu}\,\extd x^\mu\extd x^\nu = \mathcal{M}\extd u^2-2\extd u\extd r+2\mathcal{N}\extd u\extd\varphi + r^2\extd\varphi^2\,.
}{eq:cp13}
This is the same result as in Einstein gravity \cite{Barnich:2012aw}.
Exploiting the Grassmann-structure there is a neat way to define the metric again as a trace. Namely, take the matrix representation of the generators $L_n$, $M_n$, $U_n$ and $V_n$ (see appendix \ref{app:A.1}) and the twisted trace definition \eqref{eq:app13}. Then the metric is equivalently defined by
\eq{
g_{\mu\nu} = \frac{1}{2}\,\widetilde\tr \big(A_\mu A_\nu\big)\,.
}{eq:cp14}
Only bilinear expressions in the odd generators contribute to the line-element, which is precisely the statement of \eqref{eq:cp12} or \eqref{eq:cp14}.

The spin-3 field is similarly defined from the cubic sl$(3)$-Casimir or, equivalently, by using again the twisted trace 
\eq{
\Phi_{\mu\nu\lambda} = \frac{1}{6}\,\widetilde\tr\,\big(A_\mu A_\nu A_\lambda \big)
}{eq:cp15}
which for the connection \eqref{eq:cp4}-\eqref{eq:cp10} simplifies to
\eq{
\Phi_{\mu\nu\lambda} \extd x^\mu \extd x^\nu \extd x^\lambda = 2{\mathcal V}\,\extd u^3 + 4\,{\mathcal Z}\,\extd u^2\extd\varphi \,.
}{eq:cp16}
Only expressions trilinear in the odd generators contribute to the spin-3 field. 

\section{Flat space higher spin gravity with chemical potentials}\label{se:3}

\newcommand{\chemM}{\mu_{\textrm{\tiny M}}}
\newcommand{\chemV}{\mu_{\textrm{\tiny V}}}
\newcommand{\chemL}{\mu_{\textrm{\tiny L}}}
\newcommand{\chemU}{\mu_{\textrm{\tiny U}}}

In this section we generalize the discussion to flat space spin-3 gravity with chemical potentials $\chemM$, $\chemL$, $\chemV$, $\chemU$ for the spin-2 and spin-3 fields. We start by stating our main result in section \ref{se:3.1} and perform consistency checks in section \ref{se:3.2}. In section \ref{se:3.3} we discuss the canonical charges and variational principle in the presence of chemical potentials. Finally, we display results for the metric and the spin-3 field in section \ref{se:3.4}.

\subsection{Statement of the main result}\label{se:3.1}

To include chemical potentials we solve the equations of motion \eqref{eq:cp8} assuming the representation of the connection as in \eqref{eq:cp4}-\eqref{eq:cp6}. Following the procedure of \cite{Bunster:2014mua} we also assume that
the form of $a_\varphi$ remains unchanged by chemical potentials, in order to maintain the structure of the canonical charges.  We obtain
\eq{
a_u = a_u^{(0)} + a_u^{(\chemM)}  + a_u^{(\chemL)}  + a_u^{(\chemV)}  + a_u^{(\chemU)} \qquad a_\varphi = a_\varphi^{(0)}
}{eq:cp17}
with $a_u^{(0)}$, $a_\varphi^{(0)}$ being the connection \eqref{eq:cp7} in the absence of chemical potentials and 
\begin{subequations}
 \label{eq:cp18}
\begin{align}
 a_u^{(\chemM)} &= \chemM\, M_+ -\chemM^\prime\, M_0 + \tfrac12\,\big(\chemM'' - \tfrac12{\cal M} \chemM \big)\, M_- + \tfrac12\,{\cal V}\, \chemM\, V_{-2} \\
 a_u^{(\chemL)} &= a_u^{(\chemM)}\big|_{M\to L}  - \tfrac12\, {\cal N}\,\chemL\, M_- + {\cal Z}\,\chemL\, V_{-2}\\
 a_u^{(\chemV)} &= \chemV \,V_2 - \chemV^\prime\, V_1   + \tfrac12 \,\big(\chemV'' - {\cal M} \chemV\big)\, V_0 + \tfrac16\,\big(- \chemV'''+ {\cal M}^\prime \chemV  + \tfrac{5}{2} {\cal M} \chemV^\prime\big)\, V_{-1}   \nonumber \\
&\quad  + \tfrac{1}{24}\,\big(\chemV''''  - 4 {\cal M} \chemV'' - \tfrac72{\cal M}^\prime \chemV^\prime  + \tfrac32 {\cal M}^2 \chemV - {\cal M}'' \chemV \big)\,V_{-2} - 4{\cal V}\, \chemV\, M_- \\
 a_u^{(\chemU)} &= a_u^{(\chemV)}\big|_{M\to L}  - 8{\cal Z}\,\chemU\, M_-  - {\cal N}\,\chemU\, V_0 + \big(\tfrac56 {\cal N} \chemU^\prime+\tfrac13{\cal N}^\prime \chemU \big)\, V_{-1}  \nonumber \\
& \quad + \big(- \tfrac13{\cal N} \chemU'' - \tfrac{7}{24} {\cal N}^\prime \chemU^\prime   - \tfrac{1}{12}{\cal N}'' \chemU + \tfrac14 {\cal M} {\cal N} \chemU \big)\, V_{-2}  
\end{align}
where the subscript $M\to L$ denotes that in the corresponding quantity all odd generators and chemical potentials are replaced by corresponding even ones, $M_n\to L_n$, $V_n\to U_n$, $\chemM\to\chemL$ and $\chemV\to\chemU$, i.e.
\begin{align}
 a_u^{(\chemM)}\big|_{M\to L} &= \chemL\, L_+ -\chemL^\prime\, L_0 + \tfrac12\,\big(\chemL''-\tfrac12{\cal M} \chemL \big) \,L_- + \tfrac12\,{\cal V} \,\chemL\, U_{-2} \\
 a_u^{(\chemV)}\big|_{M\to L} &=  \chemU\, U_2 - \chemU^\prime\, U_1   + \tfrac12 \,\big(\chemU'' - {\cal M} \chemU\big)\, U_0 + \tfrac16\,\big(- \chemU'''+ {\cal M}^\prime \chemU  + \tfrac{5}{2} {\cal M} \chemU^\prime\big)\, U_{-1}   \nonumber \\
&\quad  + \tfrac{1}{24}\,\big(\chemU''''  - 4 {\cal M} \chemU'' - \tfrac72{\cal M}^\prime \chemU^\prime  + \tfrac32 {\cal M}^2 \chemU - {\cal M}'' \chemU \big)\,U_{-2} - 4{\cal V} \,\chemU\, L_-
\end{align}
\end{subequations}
As before, dots (primes) denote derivatives with respect to retarded time $u$ (angular coordinate $\varphi$). 

The equations of motion \eqref{eq:cp8} impose the conditions
\begin{subequations}
 \label{eq:cp19}
\begin{align}
 \dot{\cal M} &= - 2 \chemL''' + 2 {\cal M} \chemL^\prime +  {\cal M}^\prime \chemL  + 24 {\cal V} \chemU^\prime + 16 {\cal V}^\prime \chemU\label{eq:cp19a}\\
 \dot{\cal N} &=  \tfrac12\,\dot{\cal M}\big|_{L\to M} + 2 {\cal N} \chemL^\prime + {\cal N}^\prime \chemL + 24 {\cal Z} \chemU^\prime  + 16 {\cal Z}^\prime \chemU  \label{eq:cp19b}\\
 \dot{\cal V} &= \tfrac{1}{12}\,\chemU''''' - \tfrac{5}{12}\,{\cal M} \chemU''' - \tfrac58\, {\cal M}^\prime \chemU'' - \tfrac38\,{\cal M}'' \chemU^\prime  + \tfrac13\,{\cal M}^2 \chemU^\prime \nonumber \\
&\quad  - \tfrac{1}{12}{\cal M}''' \chemU  + \tfrac13\,{\cal M} {\cal M}^\prime \chemU + 3 {\cal V} \chemL^\prime  + {\cal V}^\prime \chemL  \label{eq:cp19c}\\
 \dot{\cal Z} &=   \tfrac12\,\dot{\cal V}\big|_{L\to M} - \tfrac{5}{12}\, {\cal N} \chemU''' - \tfrac58\,{\cal N}^\prime \chemU''  - \tfrac38\,{\cal N}'' \chemU^\prime + \tfrac23\, {\cal M} {\cal N} \chemU^\prime  \nonumber \\
&\quad  - \tfrac{1}{12}\,{\cal N}''' \chemU + \tfrac13\,({\cal M} {\cal N})^\prime \chemU  + 3 {\cal Z} \chemL^\prime  + {\cal Z}^\prime \chemL \label{eq:cp19d}
\end{align}
with the inverse substitution rules to above, viz.
\begin{align}
  \tfrac12\,\dot{\cal M}\big|_{L\to M} &= -\chemM''' + {\cal M} \chemM^\prime  + \tfrac12\,{\cal M}^\prime \chemM + 12 {\cal V} \chemV^\prime + 8 {\cal V}^\prime \chemV \\
 \tfrac12\,\dot{\cal V}\big|_{L\to M} &=  \tfrac{1}{24}\,\chemV''''' - \tfrac{5}{24}\,{\cal M} \chemV''' - \tfrac{5}{16}\,{\cal M}^\prime \chemV'' - \tfrac{3}{16}\,{\cal M}'' \chemV^\prime  + \tfrac16\,{\cal M}^2 \chemV^\prime  \nonumber \\
&\quad - \tfrac{1}{24}\,{\cal M}''' \chemV + \tfrac16\,{\cal M} {\cal M}^\prime \chemV + \tfrac32\,{\cal V} \chemM^\prime + \tfrac12\,{\cal V}^\prime \chemM
\end{align}
\end{subequations}
The chemical potentials $\chemM$, $\chemL$, $\chemV$ and $\chemU$ are arbitrary functions of the angular coordinate $\varphi$ and the retarded time $u$. In many applications they are constant so that many formulas simplify.

In the next subsection we provide several checks on the correctness of the results presented above and discuss in a bit more detail how we obtained them. 

\subsection{Checks}\label{se:3.2}

Note first that in the absence of chemical potentials, $\chemM=\chemL=\chemV=\chemU=0$, corresponding results from section \ref{se:2} are recovered. In particular, the on-shell conditions \eqref{eq:cp19} simplify to \eqref{eq:cp9}.
In the presence of chemical potentials the on-shell conditions \eqref{eq:cp19} contain information about the asymptotic symmetry algebra \eqref{eq:FSHSG10}. For example, the $\chemL$-terms in \eqref{eq:cp19a} are an infinitesimal Schwarzian derivative, while the $\chemU$-terms exhibit transformation behavior of a spin-3 field.
Since any solution to the field equations must be locally pure gauge, and any solution that obeys our boundary conditions can be generated by the boundary condition preserving gauge transformations \eqref{eq:cp24}, it should be possible to obtain \eqref{eq:cp18} directly from a gauge transformation. Indeed, comparing the expressions for \eqref{eq:cp24} with the expressions in \eqref{eq:cp18} we see that they coincide upon identifying $\epsilon\to\chemL$, $\tau\to\chemM$, $\kappa\to\chemV$ and $\chi\to\chemU$. This comparison provides an independent check on the correctness of our results.

It is possible to derive the results of section \ref{se:3.1} in various ways. For instance, one can start from equation (3.7)-(3.12) in \cite{Bunster:2014mua} and use the Grassmann-approach of \cite{Krishnan:2013wta} to derive the flat space connection with chemical potentials, dropping in the end all terms quadratic in the Grassmann-parameter. This is the procedure we have used. The map that leads from (3.7)-(3.12) in \cite{Bunster:2014mua} (left hand side) to the results presented in section \ref{se:3.1} (right hand side) is given by
\begin{subequations}
 \label{eq:cp60}
\begin{align}
& \textrm{coordinates:} & x^\pm &= \epsilon\,u \pm\varphi \\
& \textrm{connection\;1-form:} & 2a_\pm(x^+,x^-) &= a_u(u,\varphi)/\epsilon \pm a_\varphi(u,\varphi) \\
& \textrm{spin-2\;generators:} & 2L^\pm_n &= L_n \pm M_n/\epsilon\\
& \textrm{spin-3\;generators:} & 2W^\pm_n &= U_n \pm V_n/\epsilon\\
& \textrm{state-dependent\;spin-2\;functions:} &  \frac{24}{c_\pm}\, {\cal L}^\pm(x^\pm) &= {\cal M}(u,\varphi) \pm 2\epsilon{\cal N}(u,\varphi)\\
& \textrm{state-dependent\;spin-3\;functions:} &  -\frac{3}{c_\pm}\, {\cal W}^\pm(x^\pm) &= {\cal V}(u,\varphi) \pm 2\epsilon{\cal Z}(u,\varphi)\\
& \textrm{spin-2\;chemical\;potentials} & \tfrac14\,\xi^\pm(x^+,x^-) &= 1+\chemM(u,\varphi) \pm \chemL(u,\varphi)/\epsilon\\
& \textrm{spin-3\;chemical\;potentials} & \tfrac14\,\eta^\pm(x^+,x^-) &= \chemV(u,\varphi) \pm \chemU(u,\varphi)/\epsilon 
\end{align}
\end{subequations}
After using the map \eqref{eq:cp60} one is supposed to drop all terms quadratic (or higher power) in the Grassmann parameter $\epsilon$. Note that no inverse powers of $\epsilon$ appear anywhere in the connection, despite of their appearance in various expressions above.

Equivalently, one can do a straightforward \.In\"on\"u--Wigner contraction, sending the AdS radius to infinity. Alternatively, one could directly solve the flat space field equations \eqref{eq:cp8} with the condition that $a_\varphi$ remains unchanged as given in \eqref{eq:cp7} and only $a_u$ obtains contributions from chemical potentials. All these procedures lead to the same results displayed above in section \ref{se:3.1}.

\subsection{Canonical charges with chemical potentials}\label{se:3.3}

Since the canonical currents \eqref{eq:cp22} only depend on $a_\varphi$, which has not changed by introducing chemical potentials, the results for the canonical charges remain unchanged and all expressions displayed in section \ref{se:2.3} also apply to the case of non-vanishing $\chemM$, $\chemL$, $\chemV$ and $\chemU$. In fact, this property was the very reason why we allowed only a deformation of $a_u$. In particular, from \eqref{eq:cp26} we have the following four zero-mode charges.
\eq{
{\cal Q}_{\cal M} = \frac k2\,{\cal M}\qquad {\cal Q}_{\cal L} = k\,{\cal L}\qquad {\cal Q}_{\cal V} = 4k\,{\cal V}\qquad {\cal Q}_{\cal U} = 8k\,{\cal U}
}{eq:Q}
They can be interpreted, respectively, as mass, angular momentum, odd and even spin-3 charges.

The canonical charges will be important for our later discussion of entropy in section \ref{se:4.2} below. They also feature prominently in the variational principle. To determine the boundary term required for a well-defined variational principle we 
vary first the bulk action \eqref{eq:cp1}.
\eq{
\de I[A] = \textrm{\small bulk} + \frac{k}{4\pi}\int \langle A\wedge \delta A \rangle 
}{eq:cp83}
Evaluating the boundary term explicitly yields ($\simeq$ denotes equality up to total $\varphi$-derivative terms, which vanish upon integration over the $\varphi$-cycle)
\eq{
\langle A_\varphi \delta A_u  - A_u \delta A_\varphi \rangle \simeq {\cal M} \delta\chemM + 2{\cal N}\delta\chemL + 12{\cal V}\delta\chemV + 24{\cal Z}\delta\chemU + 4\chemV\delta{\cal V} + 8\chemU\delta{\cal Z}\,.
}{eq:cp84}
This confirms the result \cite{Afshar:2013vka} that the bulk action \eqref{eq:cp1} has a well-defined variational principle in the absence of spin-3 chemical potentials. In their presence, however, the last two terms are incompatible with a well-defined variational principle.  Therefore, we subtract a boundary counterterm to restore a well-defined variational principle for this case,
\eq{
\Gamma[A] = I[A] - I_b[A]\qquad \textrm{with}\qquad I_b[A] = \frac{k}{4\pi} \int\!\extd u\extd\varphi\,  \langle \bar A_u A_\varphi \rangle 
}{eq:cp85}
where $\bar A_u = b^{-1} \bar a_u b$ with the same group element $b$ as before [see Eq.~\eqref{eq:cp5}] and
\eq{
\bar a_u = a_u - 2 (1+\chemM)\, M_+ - 2\chemL\,L_+ - 2\chemV\,V_2 - 2\chemU\,U_2\,.
}{eq:cp86}
In total we get (${\cal Q}_{\cal N}$ is ${\cal Q}_{\cal L}$ with $\cal L$ replaced by $\cal N$, and similarly for ${\cal Q}_{\cal Z}$ with ${\cal Q}_{\cal U}$)
\begin{multline}
\delta\Gamma\big|_{\textrm{\tiny EOM}} =  \frac{k}{4\pi} \int\!\extd u\extd\varphi\, \big(\langle A_\varphi \delta A_u  - A_u \delta A_\varphi \rangle - \delta \langle \bar A_u A_\varphi \rangle\big) \\
= \int\extd u \,\big({\cal Q}_{\cal M}\,\de\chemM + {\cal Q}_{\cal N}\,\de\chemL + {\cal Q}_{\cal V}\,\de\chemV + {\cal Q}_{\cal Z}\,\de\chemU\big)\,.
\label{eq:cp87}
\end{multline}
In conclusion, the action \eqref{eq:cp85} has a well-defined variational principle, in the sense that the first variation of the full action vanishes on-shell for arbitrary (but fixed) chemical potentials. As expected, the response functions \eqref{eq:cp87} are determined by  the canonical charges, and the chemical potentials act as sources.

\subsection{Metric and spin-3 field in presence of chemical potentials}\label{se:3.4}

Plugging the results for the connection with chemical potentials, \eqref{eq:cp17}-\eqref{eq:cp19} with \eqref{eq:cp4}-\eqref{eq:cp6}, into the definitions for the metric \eqref{eq:cp14} yields 
\eq{
g_{\mu\nu}\,\extd x^\mu\extd x^\nu = g_{uu}\extd u^2 + g_{u\varphi}\,2\extd u\extd\varphi - (1+\chemM)\,2\extd r\extd u  + r^2\extd\varphi^2
}{eq:cp20}
with
\begin{subequations}
\label{eq:cp20a}
\begin{align}
g_{uu} &=   r^2\,\big(\chemL^2  - 4 \chemU''\chemU + 3 \chemU^{\prime\,2} + 4 {\cal M} \chemU^2\big) 
+ r\,g_{uu}^{(r)} + g_{uu}^{(0)} + g_{uu}^{(0^\prime)}\label{eq:cp20aa} \\
g_{u\varphi} &= r^2\chemL - r\chemM^\prime  + {\cal N}(1+\chemM)  + 8{\cal Z}\chemV \label{eq:cp20ab}
\end{align}
where
\begin{align}
g_{uu}^{(0)} &=   {\cal M} (1+\chemM)^2   + 2(1+\chemM)\big({\cal N} \chemL + 12 {\cal V} \chemV + 16 {\cal Z} \chemU\big) \nonumber \\
&\quad + 16 {\cal Z} \chemL \chemV  + \tfrac43 \big({\cal M}^2 \chemV^2 + 4{\cal M} {\cal N} \chemU \chemV + {\cal N}^2 \chemU^2\big)  
\end{align}
\end{subequations}
and the contributions $g_{uu}^{(r)}$ and $g_{uu}^{(0^\prime)}$ are presented in \eqref{eq:app42} in appendix \ref{app:B}.

Similarly, we obtain from the definition of the spin-3 field \eqref{eq:cp16}
\begin{multline}
\!\!\!\!\!\!\Phi_{\mu\nu\lambda} \extd x^\mu \extd x^\nu \extd x^\lambda = \Phi_{uuu} \extd u^3+\Phi_{ruu} \extd r \extd u^2+\Phi_{uu\varphi} \extd u^2 \extd\varphi  - \big(2\chemU r^2 - r\chemV^\prime + 2{\cal N}\chemV\big)\!\extd r\extd u \extd\varphi \\ 
+ \chemV \extd r^2\extd u -  \big(\chemU^\prime r^3 - \tfrac13r^2(\chemV''-{\cal M}\chemV + 4 {\cal N} \chemU) + r{\cal N}\chemV^\prime - {\cal N}^2\chemV \big) \extd u \extd\varphi^2 
\label{eq:cp21}
\end{multline} 
with
\begin{subequations}
 \label{eq:cp28}
\begin{align}
 \Phi_{uuu} &= r^2\,\big[2 (1+\chemM) \chemU ({\cal M} \chemL  - 4 {\cal V}\chemU) - \tfrac13\chemL^2  ( {\cal M} \chemV - 4 {\cal N} \chemU )  \nonumber \\
&\qquad \quad   + 16\chemL \chemU( {\cal V} \chemV  + {\cal Z} \chemU ) - \tfrac43 {\cal M} \chemU^2 ({\cal M}\chemV + 2 {\cal N} \chemU)    \big] \nonumber \\
&\quad  + 2{\cal V}(1+\chemM)^3  +  \tfrac23 (1+\chemM)^2 \big(6 {\cal Z}\chemL + {\cal M}^2 \chemV + 2 {\cal M} {\cal N} \chemU\big)  \nonumber \\
&\quad  + \tfrac23 (1+\chemM) \big( ({\cal N} \chemL + 16{\cal Z}\chemU) ( 2{\cal M} \chemV + {\cal N} \chemU) + 12 {\cal M} {\cal V} \chemV^2  \big)  + 
 {\cal N}^2 \chemL^2 \chemV \nonumber \\
&\quad  + 16  \chemL \chemV^2 ({\cal N} {\cal V} - 
 \tfrac{1}{3} {\cal M} {\cal Z}) + \tfrac{64}{3} {\cal Z} \chemU \chemV ({\cal N} \chemL + 12{\cal V}\chemV + 12{\cal Z} \chemU )  + 64 {\cal V}^2  \chemV^3 \nonumber \\
&\quad - \tfrac{8}{27} ({\cal M}^3 \chemV^3  - {\cal N}^3 \chemU^3) - \tfrac49 {\cal M} {\cal N} \chemU \chemV (4{\cal M} \chemV + 5 {\cal N} \chemU) \nonumber \\
&\quad + r^3\,\Phi_{uuu}^{(r^3)} + r^2\,\Phi_{uuu}^{(r^2)} + r\,\Phi_{uuu}^{(r)} + \Phi_{uuu}^{(0)} \label{eq:cp28a}  \displaybreak[1] \\
 \Phi_{ruu} &= - 2r^2 \chemL \chemU   - \tfrac23(1+\chemM)(2{\cal M}\chemV  + {\cal N} \chemU) - 2 {\cal N} \chemL \chemV  \nonumber \\
&\quad  - 16\chemV ({\cal V}\chemV + 2 {\cal Z} \chemU) + r \, \Phi_{ruu}^{(r)}  +  \Phi_{ruu}^{(0)}   \displaybreak[1] \\
 \Phi_{uu\varphi} &= r^2\,\big[2 {\cal M} (1+\chemM) \chemU   - \tfrac23 \chemL ({\cal M} \chemV - 4{\cal N} \chemU) + 16\chemU ({\cal V} \chemV + {\cal Z} \chemU) \big] \nonumber \\
&\quad  + 4{\cal Z}(1\!+\!\chemM)^2  + \tfrac23 {\cal N} (1\!+\!\chemM) (2{\cal M}\chemV + {\cal N} \chemU) + 2{\cal N}\chemV({\cal N}\chemL  + \tfrac{32}{3} {\cal Z} \chemU) \nonumber \\
&\quad - \tfrac{16}{3}({\cal M}{\cal Z}-3{\cal V}{\cal N})\chemV^2  + r^3\,\Phi_{uu\varphi}^{(r^3)} + r^2\,\Phi_{uu\varphi}^{(r^2)} + r\,\Phi_{uu\varphi}^{(r)}  + \Phi_{uu\varphi}^{(0)} 
\end{align}
\end{subequations}
where the contributions $\Phi_{uuu}^{(r^3)}$, $\Phi_{uuu}^{(r^2)}$, $\Phi_{uuu}^{(r)}$, $\Phi_{uuu}^{(0)}$, $\Phi_{ruu}^{(r)}$, $\Phi_{ruu}^{(0)}$, $\Phi_{uu\varphi}^{(r^3)}$, $\Phi_{uu\varphi}^{(r^2)}$, $\Phi_{uu\varphi}^{(r)}$ and $\Phi_{uu\varphi}^{(0)}$ are collected in appendix \ref{app:B}.

Note that for zero-mode solutions with constant chemical potentials, ${\cal M}^\prime={\cal N}^\prime=\chemM^\prime=\chemL^\prime=\chemV^\prime=\chemU^\prime=0$, all the expressions in appendix \ref{app:B} vanish and thus the spin-2 and spin-3 fields simplify considerably in this case (see also appendix \ref{se:4.1}).

\section{Flat space Einstein gravity with chemical potentials}\label{sec:E}

If we set to zero the spin-3 charges and spin-3 chemical potentials, ${\cal V}={\cal Z}=\chemV=\chemU=0$, we recover flat space Einstein gravity with chemical potentials $\chemM$ and $\chemL$. While this is merely a special case of the more general results of section \ref{se:3}, it seems convenient for future applications to collect these results separately and to elaborate on them. This is what we do in this section.

In section \ref{se:E.1} we present the general solution for the isl$(2)$ gauge connection and the metric with arbitrary spin-2 chemical potentials. In section \ref{se:E.2} we focus on zero mode solutions with constant chemical potentials and provide a canonical interpretation of the latter. In section \ref{se:E.3} we linearize the solutions in the chemical potentials, which is useful for some applications, like the holographic dictionary, which we address in section \ref{se:E.4}.

\subsection{General solution}\label{se:E.1}

The connection is given by \eqref{eq:cp4}, \eqref{eq:cp5}, \eqref{eq:cp6} with
\begin{subequations}
 \label{eq:cp50}
\begin{align}
 a_u &= (1+\chemM)\,M_+ - \chemM^\prime\,M_0 + \tfrac12\,\big(\chemM''-\tfrac12{\cal M}(1+\chemM) - {\cal N}\,\chemL \big)\,M_- \nonumber \\
&\quad + \chemL\, L_+ -\chemL^\prime\, L_0 + \tfrac12\,\big(\chemL''-\tfrac12{\cal M} \chemL \big) \,L_- \\
 a_\varphi &= L_+ - \frac{\cal M}{4}\,L_- - \frac{\cal N}{2}\,M_-\,.
\end{align}
\end{subequations}
The corresponding line-element reads
\begin{multline}
g_{\mu\nu}\,\extd x^\mu\extd x^\nu = \big[r^2\chemL^2 + 2 r \big(\chemL^\prime (1+\chemM) - \chemL \chemM^\prime\big) +  {\cal M} (1+\chemM)^2   + 2(1+\chemM)({\cal N} \chemL - \chemM'') \\
+\chemM^{\prime\,2} \big]\extd u^2 
+ \big(r^2\chemL - r\chemM^\prime  + {\cal N}(1+\chemM)  \big)\,2\extd u\extd\varphi - (1+\chemM)\,2\extd r\extd u  + r^2\extd\varphi^2
\label{eq:cp49}
\end{multline}
with the on-shell conditions
\begin{subequations}
\label{eq:cp53}
 \begin{align}
 \dot{\cal M} &= - 2 \chemL''' + 2 {\cal M} \chemL^\prime + {\cal M}^\prime \chemL \label{eq:cp53a}\\
 \dot{\cal N} &= - \chemM''' + {\cal M} \chemM^\prime + \tfrac12{\cal M}^\prime \chemM + 2 {\cal N} \chemL^\prime + {\cal N}^\prime \chemL \,. \label{eq:cp53b}
 \end{align}
\end{subequations}

\subsection{Zero mode solutions with constant chemical potentials}\label{se:E.2}

We consider now zero mode solutions, ${\cal M}^\prime={\cal N}^\prime=0$, with constant spin-2 chemical potential, $\chemM^\prime=\chemL^\prime=0$. Then the results above simplify further.
The line-element reads
\begin{multline}
g_{\mu\nu}\,\extd x^\mu\extd x^\nu = \big[r^2\chemL^2 +  {\cal M} (1+\chemM)^2   + 2{\cal N}(1+\chemM) \chemL  \big]\extd u^2 \\
+ \big(r^2\chemL + {\cal N}(1+\chemM)  \big)\,2\extd u\extd\varphi - (1+\chemM)\,2\extd r\extd u  + r^2\extd\varphi^2
\label{eq:cp61}
\end{multline}
with the on-shell conditions $\dot{\cal M}=\dot{\cal N}=0$. 

If we set to zero the even chemical potential, $\chemL=0$, then the line-element \eqref{eq:cp61} simplifies to the vacuum solution \eqref{eq:cp11}, but with $u$ replaced by $\tilde u = (1+\chemM) u$. Therefore, a constant odd chemical potential $\chemM$ effectively rescales the retarded time coordinate. In canonical general relativity language the odd chemical potential $\chemM$ rescales the lapse function.

If instead we set to zero the odd chemical potential, $\chemM=0$, then the line-element \eqref{eq:cp61} simplifies to
\begin{align}
g_{\mu\nu}\,\extd x^\mu\extd x^\nu &= \Big({\cal M} - \frac{\cal N}{r^2}  \Big)\extd u^2 - 2\extd r\extd u  + r^2\,\Big(\extd\varphi + \frac{\cal N}{r^2}\extd u + \chemL \extd u\Big)^2\,.
\label{eq:cp70} \\
\intertext{Comparing this result with the vacuum solution \eqref{eq:cp11} in ADM-like form,}
g_{\mu\nu}\,\extd x^\mu\extd x^\nu &= \Big({\cal M} - \frac{\cal N}{r^2}  \Big)\extd u^2 - 2\extd r\extd u  + r^2\,\Big(\extd\varphi + \frac{\cal N}{r^2}\extd u \Big)^2
\label{eq:cp71}
\end{align}
we see that the even chemical potential $\chemL$ changes only the last term. In canonical general relativity language the even chemical potential $\chemL$ shifts the shift vector.

\subsection{Perturbative solutions linearized in chemical potentials}\label{se:E.3}

A different kind of simplification arises when linearizing in the chemical potentials. Expanding the metric \eqref{eq:cp20} in the chemical potentials,
\eq{
g_{\mu\nu} = \bar g_{\mu\nu} + h_{\mu\nu} + {\cal O}(\chemM^2,\chemL^2,\chemM\chemL)
}{eq:cp65}
with the background line-element $\bar g_{\mu\nu}\extd x^\mu\extd x^\nu$ given by the right hand side of \eqref{eq:cp13}, yields for the linear terms
\begin{multline}
h_{\mu\nu} \extd x^\mu \extd x^\nu =  2\big( {\cal M}\, \chemM + {\cal N}\, \chemL\big) \extd u^2 + \big(r^2\,\chemL + {\cal N}\, \chemM \big) 2\extd u \extd\varphi - 2 \chemM \,\extd r \extd u  \\
+ 2\big(r\, \chemL^\prime - \chemM''\big)\extd u^2 - 2r\, \chemM^\prime\,\extd u\extd\varphi \,.
 \label{eq:cp66}
\end{multline}
The terms in the second line vanish for constant chemical potentials.

\subsection{Comparison with holographic dictionary}\label{se:E.4}

From a holographic perspective, the first two terms in the linearized solution \eqref{eq:cp66} show the typical coupling between sources (chemical potentials) and vacuum expectation values (canonical charges). The $r^2\chemL \extd u\extd\varphi$ term and the $\chemM\extd r\extd u$ term correspond to the essential terms in the two towers of non-normalizable\footnote{%
Here and in what follows the attribute ``non-normalizable'' always means ``breaking the Barnich--Comp{\`e}re boundary conditions'' \cite{Barnich:2006av} or the corresponding spin-3 version \cite{Afshar:2013vka,Gonzalez:2013oaa}. 
}  solutions to the linearized equations of motion. 

In the holographic dictionary, these non-normalizable contributions should correspond to sources of the corresponding operators in the dual field theory. Indeed, this is what happens as shown in \cite{Detournay:2014fva}. Note, however, that \cite{Detournay:2014fva} worked in Euclidean signature, restricted to zero mode solutions and imposed axial gauge for the non-normalizable solutions to the linearized Einstein equations on a flat space background, so a direct comparison is not straightforward. Exploiting our interpretation of constant chemical potentials as modifications of lapse and shift (see section \ref{se:E.2}) we can interpret the results of \cite{Detournay:2014fva} as follows (see their section 3.4): their quantity $\delta \xi_J$ corresponds precisely to the (linearized) even chemical potential $\delta\xi_J\sim\chemL$, and their quantity $\delta \xi_M$ corresponds to twice the (linearized) odd chemical potential, $\delta \xi_M\sim 2\chemM$. This identification is perfectly consistent with the 
holographic interpretation summarized above.

\section{Applications}\label{se:4}

In this section we address some applications, without claiming to be exhaustive. 

 In section \ref{se:4.2} we calculate the entropy of flat space cosmologies with spin-3 charges by solving all holonomy conditions. In section \ref{se:4.new} we determine the free energy and discover novel types of phase transitions.  In section \ref{se:4.4} we conclude with some remarks on the recent spin-3 singularity resolution of flat space orbifolds. In section \ref{se:4.5} we provide an outlook to some further possible applications. 

As supplements, in appendix \ref{se:4.1} we discuss zero-mode solutions with constant spin-3 and vanishing spin-2 chemical potentials and in appendix \ref{se:4.3} we consider more general solutions to the field equations, dubbed ``chemically odd'', by restricting to odd chemical potentials only and by allowing specific deformations of $a_\varphi$.

\subsection{Entropy}\label{se:4.2}

\newcommand{\odd}{v}
\newcommand{\coL}{f_L}
\newcommand{\coU}{f_U}
\newcommand{\defP}{\mu}
\newcommand{\defR}{\eta}
\newcommand{\defT}{\theta}

To discuss thermodynamical aspects we restrict ourselves to zero mode solutions with constant chemical potentials.
The main quantity of interest is the entropy of solutions like flat space cosmologies with spin-3 charges switched on.
As we shall demonstrate by solving holonomy conditions, entropy is given by a hatted trace,
\eq{
S = 
2k \beta_L\,\widehat\tr\big(a_u a_\varphi\big)\Big|_{\textrm{EOM}} = 
\beta_L\,\big( 2(1 + \chemM) {\cal Q}_{\cal M}  + 2\chemL {\cal Q}_{\cal L} + 3 \chemV {\cal Q}_{\cal V} + 3 \chemU {\cal Q}_{\cal U}\big)\,.
}{eq:S}
The quantity $\beta_L$ is not necessarily the inverse temperature, but rather the length of the relevant cycle appearing in the holonomy condition below. The zero mode charges ${\cal Q}_i$ are displayed in \eqref{eq:Q}.

We start by proposing the holonomy condition that we want to solve.
\eq{
\exp{\big(i\beta_L a_u\big)} = \unity
}{eq:cp100}
This condition is completely analogous to corresponding holonomy conditions for higher spin black holes in AdS \cite{Gutperle:2011kf}. To solve the holonomy condition \eqref{eq:cp100} we exploit the representation summarized in appendix \ref{app:A.2} in terms of $9\times 9$ matrices. By a similarity transformation we can diagonalize the ad-part of a generic matrix of the form \eqref{eq:ad1}. 
\eq{
\begin{pmatrix}
 A^{-1}_{8\times 8} & \mathbb{O}_{8\times 1} \\
 \mathbb{O}_{1\times 8} & 1
\end{pmatrix} 
\begin{pmatrix}
     \textrm{ad}_{8\times 8} & \textrm{odd}_{8\times 1} \\
     \mathbb{O}_{1\times 8} & 0
\end{pmatrix}
\begin{pmatrix}
 A_{8\times 8} & \mathbb{O}_{8\times 1} \\
 \mathbb{O}_{1\times 8} & 1
\end{pmatrix} =
\begin{pmatrix}
     \big(A^{-1}\textrm{ad}A\big)_{8\times 8} & \big(A^{-1}\textrm{odd}\big)_{8\times 1} \\
     \mathbb{O}_{1\times 8} & 0
\end{pmatrix}
}{eq:cp101}
A matrix of this form is easily exponentiated. Assuming that ad has zero as eigenvalue with geometric and algebraic multiplicity $n$ and denoting $\odd=A^{-1}\textrm{odd}$ yields
\eq{
\exp{\begin{pmatrix}
     \big(A^{-1}\textrm{ad}A\big)_{8\times 8} & \big(A^{-1}\textrm{odd}\big)_{8\times 1} \\
     \mathbb{O}_{1\times 8} & 0
     \end{pmatrix}
} = \left(\begin{array}{ccccccl}
1 \quad &&&&&                 & \odd_1 \\
  & \ddots &&&&               & \vdots \\
  && \quad 1 \; &&&           & \odd_n \\
  &&& \; e^{\lambda_1}\; &&   & \odd_{n+1} \frac{e^{\lambda_1}-1}{\lambda_1} \\
  &&&& \ddots &               & \vdots \\
  &&&&&\;\,e^{\lambda_{8-n}}\;& \odd_8 \frac{e^{\lambda_{8-n}}-1}{\lambda_{8-n}} \\
  &&&&&                       & 1
    \end{array}\right)\,.
}{eq:cp102}
In our case $n=2$ [$=\textrm{rank\;sl}(3)$] and the holonomy condition \eqref{eq:cp100} is then solved by the relations
\eq{
\lambda_k = 0 \;\textrm{mod}\; \frac{2\pi}{\beta_L} \,, \quad k=1..6\,; \qquad \odd_m = 0 \,,\quad m=1..2 \,.
}{eq:cp103}

The first set of relations \eqref{eq:cp103} is precisely the same as in AdS spin-3 gravity for one chiral half. Therefore, we must be able to represent these conditions in the same way as it was done in AdS.
In fact, a plausible guess for the two holonomy conditions that follow from the first set of relations \eqref{eq:cp103} is given by (compare with corresponding conditions in the AdS case, particularly Eqs.~(3.32) and (3.33) in \cite{Bunster:2014mua})
\begin{align}
 \tfrac14\tr\big(a_u a_u\big)\Big|_{\epsilon=0} &= {\cal M} \chemL^2 + 24 {\cal V} \chemL \chemU + \tfrac43 {\cal M}^2 \chemU^2 = \frac{4\pi^2}{\beta_L^2} \label{eq:cp106} \\
 \tfrac14\sqrt{\det a_u}\Big|_{\epsilon=0} &= \big|{\cal V} \chemL^3 + \tfrac13 {\cal M}^2 \chemL^2 \chemU + 4 {\cal M} {\cal V} \chemL \chemU^2 - \tfrac{4}{27} {\cal M}^3 \chemU^3 + 32 {\cal V}^2 \chemU^3\big| = 0  \label{eq:cp107}
\end{align}
We prove now that this is indeed the correct result.
Since the matrix $A^{-1}\textrm{ad}A$ is diagonal, it must lie in the Cartan subalgebra of sl$(3)$; diagonalizing simultaneously $L_0$ and $U_0$ we find 
\eq{
A^{-1}\textrm{ad}A = \textrm{diag}\,\big(0,0,\coL+2\coU,\coL-2\coU,-\coL+2\coU,-\coL-2\coU,2\coL,-2\coL\big)
}{eq:cp104}
with some functions $\coL$, $\coU$ of the charges and chemical potentials that can be determined by explicitly calculating the characteristic polynomial of the matrix $i\beta_L a_u$ for the eigenvalues $\lambda$ as derived from the solution \eqref{eq:cp18} (with constant charges and chemical potentials) and comparing it with the characteristic polynomial that follows from \eqref{eq:cp104}.
The first set of relations \eqref{eq:cp103} yields the conditions
\eq{
\coL = \frac{m \pi}{\beta_L} \qquad \coU = \frac{(n-\tfrac m2)\pi}{\beta_L}\qquad n,m\in\mathbb{Z}\,.
}{eq:cp105}
Thus, the first half of the holonomy conditions leads to a discrete family of solutions parametrized by two integers $n$ and $m$. For the choice $m=2$ and $n=1$ these conditions reproduce precisely the guess \eqref{eq:cp106} and \eqref{eq:cp107}. This choice is unique by requiring that in the absence of spin-3 chemical potentials and spin-3 charges the holonomy conditions reduce to the ones for flat space cosmologies. We will therefore always make this choice in the present work.

So far we have obtained and solved only half of the holonomy conditions. The other half emerges from imposing the second set of relations \eqref{eq:cp103}. After a straightforward calculation\footnote{%
There are numerous different ways to obtain these results, but it is not always easy to extract the simple conditions \eqref{eq:cp109} and \eqref{eq:cp110}. For instance, one can contract the AdS holonomy conditions using the map \eqref{eq:cp60}, but this leads naturally to non-linear relations between charges and chemical potentials. Two combinations of these relations immediately provide the holonomy conditions \eqref{eq:cp106} and \eqref{eq:cp107}, but it takes a bit of work to extract the other two conditions in their simplest form. Alternatively, one can explicitly construct the matrix $A$ in \eqref{eq:cp101} that diagonalizes the sl$(3)$ part of the generators and then determine the two eigenvectors associated with the two zero eigenvalues. This approach makes it clear from the start that the remaining two holonomy conditions must be linear in the chemical potentials. The procedure we used is a simpler version thereof that avoids complete diagonalization, but merely puts the generators into block form 
with a $2\times 2$ block of zeros, since the remaining two holonomy conditions are restricted to the subspace associated with the zero eigenvalues.
} we find that one of these conditions is linear in the charges and chemical potentials, while the other is quadratic in the charges and linear in the chemical potentials
\begin{align}
  \mathcal{M}(1+\chemM) + \mathcal{L} \chemL + 12 \mathcal{V} \chemV + 16 \mathcal{U} \chemU &= 0 \label{eq:cp109} \\
  9  \mathcal{V} (1+\chemM) + 6  \mathcal{U} \chemL + \mathcal{M}^2 \chemV + 2 \mathcal{L} \mathcal{M} \chemU  &=0\,. \label{eq:cp110}
\end{align}
These results are considerably simpler than the corresponding holonomy conditions in AdS, which are at least quadratic in chemical potentials and charges.

The linear holonomy condition \eqref{eq:cp109} simplifies entropy \eqref{eq:S} to
\eq{
S = 
\beta_L\,\big(\chemL {\cal Q}_{\cal L} + \chemU {\cal Q}_{\cal U}\big)\,.
}{eq:S2}
For the special case $\chemU=0$ entropy \eqref{eq:S2} depends only on spin-2 charges and chemical potentials (see appendices \ref{se:4.1} and \ref{se:4.3}). Moreover, the solution to the four holonomy conditions \eqref{eq:cp106}, \eqref{eq:cp107}, \eqref{eq:cp109}, \eqref{eq:cp110} becomes elementary.
\eq{
\mathcal{M} = \frac{4\pi^2}{\beta_L^2\chemL^2} \qquad \mathcal{L} = - \mathcal{M} \frac{1+\chemM}{\chemL} \qquad \mathcal{V}=0\qquad \mathcal{U} = -\mathcal{M}^2\,\frac{\chemV}{6\chemL}
}{eq:cp108}
For that case entropy is given by the Bekenstein--Hawking area law ($k=1/(4G_N)$, where $G_N$ is Newton's constant)
\eq{
S\big|_{\chemU=0} = 
k\beta_L\,|\chemL {\cal L}| = 
k\,\frac{2\pi|{\cal L}|}{\sqrt{\cal M}} = k \,\textrm{area}_{\textrm{\tiny horizon}}\,.
}{eq:S2a}
We included absolute values to ensure that entropy is positive regardless of the sign of the charge ${\cal L}$.
Inverse temperature
\eq{
\beta = -\frac{\partial S}{\partial {\cal Q}_{\cal M}}\Big|_{{\cal Q}_{\cal L}} = -\frac{2\partial S}{k \partial {\cal M}}\Big|_{\cal L} = 
2\pi \,\frac{|{\cal L}|}{{\cal M}^{3/2}}
}{eq:cp111}
then coincides with the spin-2 result (see e.g.~\cite{Bagchi:2013lma}; note that in their conventions ${\cal M}=r_+^2$ and $|{\cal L}|=|r_0r_+|$).
\eq{
T = \frac{1}{2\pi} \frac{{\cal M}^{3/2}}{|{\cal L}|} \,.
}{eq:cp73}
The minus sign in the definition \eqref{eq:cp111} is reminiscent of the inner horizon first law of black hole mechanics \cite{Larsen:1997ge, Cvetic:1997uw, Curir:1981uc, Castro:2012av, Detournay:2012ug} as explained in \cite{Bagchi:2013lma}. From the corresponding first law
\eq{
-\extd {\cal Q}_{\cal M} = T\,\extd S + \Omega\,\extd Q_{\cal L}
}{eq:cp113}
we deduce the angular potential 
\eq{
\Omega = -T\,\frac{\partial S}{\partial {\cal Q}_{\cal L}}\Big|_{{\cal Q}_{\cal M}} = -T\,\frac{\partial S}{k \partial {\cal L}}\Big|_{\cal M} = 
\frac{\cal M}{\cal L}  
}{eq:cp112}
which again coincides with the spin-2 result \cite{Bagchi:2013lma}.

In the general case $\chemU\neq 0$ not all holonomy conditions are linear. Instead, we have to solve one quadratic and one cubic equation, similar to the AdS case. Defining $\defP=\chemL\chemU$ and $\defR=\chemL/\chemU + \tfrac19 {\cal M}^2/\cal V$ the holonomy conditions \eqref{eq:cp106}, \eqref{eq:cp107} simplify to
\begin{align}
& \defR^3 + \defR\Big(4{\cal M}-\frac{{\cal M}^4}{27{\cal V}^2}\Big) + 32{\cal V} - \frac{16{\cal M}^3}{27\cal V} + \frac{2{\cal M}^6}{729{\cal V}^3} = 0 \label{eq:cp114} \\
& \defP = \frac{4\pi^2}{\beta_L^2}\,\Big(\frac{\chemL}{\chemU}{\cal M} + 24{\cal V}+\frac{4\chemU}{3\chemL}{\cal M}^2\Big)^{-1}\,. \label{eq:cp115}
\end{align}
Solving the cubic equation \eqref{eq:cp114} yields a result for the ratio $\chemL/\chemU$, which can then be plugged into the linear equation \eqref{eq:cp115} to determine the product of the chemical potentials. The sign of the discriminant $D$ of the cubic equation \eqref{eq:cp114} is given by
\eq{
\textrm{sign}\, D = \textrm{sign}\big({\cal M}^3-108{\cal V}^2\big)\,.
}{eq:cp116}
If $D$ is negative there is exactly one real solution; this happens only if the spin-3 charge $\cal V$ is sufficiently large or if the mass $\cal M$ is negative. For a critical tuning of the charges, 
\eq{
\textrm{criticality:}\qquad 108{\cal V}^2={\cal M}^3 
}{eq:cp117}
the discriminant vanishes, $D=0$, and there is a unique real solution $\defR=0$. However, the linear equation \eqref{eq:cp116} has no finite solution for $\defP$ in this case. Therefore, starting from finite and positive ${\cal M}$ it is not possible to smoothly increase the spin-3 charge ${\cal V}$ beyond the critical value \eqref{eq:cp117}. 

Heneceforth, we shall always assume the inequality
\eq{
{\cal M} > \big(108{\cal V}^2\big)^{1/3} \geq 0\,.
}{eq:cp120}
In other words, we consider from now on exclusively the case of positive discriminant, $D>0$. In this case there are three real solutions for $\defR$. 
%
The resulting entropy is real for all three branches. However, only one branch recovers the same entropy \eqref{eq:S2} as for the spin-2 case in the limit ${\cal V}\to 0$. Therefore, we take that branch.

On this particular branch, there is a neat way to express all results in terms of the charges ${\cal M,L, U}$ and a new parameter ${\cal R}$ that depends on the ratio of spin-3 and spin-2 charges ${\cal V}^2/{\cal M}^3$, just like in the AdS case \cite{Gutperle:2011kf}:
\eq{
\frac{{\cal R}-1}{4{\cal R}^{3/2}}=\frac{|{\cal V}|}{{\cal M}^{3/2}} \qquad {\cal R} > 3
}{eq:cp123}
The restriction to ${\cal R} > 3$ guarantees that we sit on the correct branch. 
The chemical potentials then read
\begin{align}
 1+\chemM &= -\frac{2 \pi|{\cal L}|}{{\cal M}\sqrt{\cal M}\beta_L}\,\cdot\,\frac{4{\cal R}(2{\cal R}^2+6{\cal R}-9) -  24{\cal P}\sqrt{\cal R}(10{\cal R}^2 - 15{\cal R} + 9) }{ ({\cal R}-3)^3 (4-3/{\cal R})^{3/2}} \\
 \chemL &= \frac{2\pi\,\textrm{sign}{\cal L}}{\sqrt{\cal M}\beta_L}\,\cdot\,\frac{2{\cal R}-3}{({\cal R}-3)\sqrt{4-3/{\cal R}}}\\
 \chemU &= -\frac{3\pi\,\textrm{sign}{\cal L}}{{\cal M}\beta_L}\,\cdot\,\frac{\sqrt{\cal R}}{({\cal R}-3)\sqrt{4-3/{\cal R}}} \\
 \chemV &= \frac{3\pi|{\cal L}|}{{\cal M}^2\beta_L}\,\cdot\,\frac{2\sqrt{\cal R}(10{\cal R}^2-15{\cal R}+9) - 16{\cal P R} (2 {\cal R}^2+6{\cal R} - 9)}{({\cal R}-3)^3(4-3/{\cal R})^{3/2}}
\end{align}
while entropy is given by
\eq{
S({\cal M,L,R,P}) = 
2\pi k \, \frac{|{\cal L}|}{\sqrt{\cal M}}\,\cdot\,\frac{2{\cal R}-3 - 12{\cal P}\sqrt{\cal R}}{({\cal R}-3)\sqrt{4-3/{\cal R}}}  \,.
}{eq:cp124}
with the dimensionless ratio \eq{
{\cal P} = \frac{{\cal U}}{\sqrt{\cal M}{\cal L}}\,.
}{eq:cp125}
The expression for entropy \eqref{eq:cp124} is the main result of this section. The pre-factor containing the spin-2 charges ${\cal M, L}$ coincides with the spin-2 result \eqref{eq:S2a}. The spin-3 correction depends non-linearly on one of the combinations of spin-3 charges, ${\cal R}$, and linearly on the other, ${\cal P}$. 

For some purposes it can be useful to have a simpler perturbative result for entropy in the limit of small spin-3 charge ${\cal V}$ (large $\cal R$), which we present below.
\eq{
S({\cal M,L,V,U}) = 
2\pi k \,\frac{|{\cal L}|}{\sqrt{\cal M}}\,\Big(1+\frac{15{\cal V}^2}{8{\cal M}^3} - \frac{6 {\cal U}|{\cal V}|}{{\cal M}^2{\cal L}} \Big) + {\cal O}({\cal V}^{3})
}{eq:cp126}

We close the entropy discussion by addressing sign issues. We have assumed that the mass is positive, ${\cal M}>0$, motivated by the necessity of this condition in the spin-2 case. The sign of $\cal L$ does not matter, which is why we included absolute values in the final result for entropy \eqref{eq:cp124}. Here is our argument. Suppose that ${\cal L}>0$ (${\cal L}<0$). Then we exploit the sign ambiguity in the definitions of $\chemL$, $\chemU$ by choosing $\chemL>0$ ($\chemL<0$) so that the first term in \eqref{eq:S2} is always positive and thus entropy is positive in the limit of vanishing spin-3 fields. The sign of ${\cal V}$ is taken care of by the definition \eqref{eq:cp123}, which ensures positive ${\cal R}$ regardless of the sign of ${\cal V}$. Thus, the only remaining signs of potential relevance are the signs of the spin-3 charge $\cal U$ and the corresponding chemical potential $\chemU$. The latter is fixed through the sign choice of $\chemL$ explained above, but the former is free to change, and 
this change is physically relevant. This implies that the quantity $\cal P$ defined in \eqref{eq:cp125} can have either sign, so that the last term in the entropy \eqref{eq:cp124} can have either sign. Demanding positivity of entropy then establishes an upper bound on ${\cal U}$.

\subsection{Grand canonical free energy and phase transitions}\label{se:4.new}

In the previous section we found that there are three branches of solutions of all the holonomy conditions, and we simply took the branch that connects continuously to the spin-2 results in the limit of vanishing spin-3 charges. However, it is not guaranteed that this procedure picks out the correct branch from a thermodynamical perspective in the whole parameter space. What we should do is to compare the free energies of all branches for given values of the chemical potentials and check which of the branches leads to the lowest free energy. This is precisely the aim of this subsection.

We start by writing the general result for the (grand canonical) free energy, regardless of the specific branch (we set $k=1$ in this subsection). We already have a thermodynamic potential, namely entropy in terms of extensive quantities (charges), so all we need to do is to Legendre transform with respect to all pairs charge/chemical potential.\footnote{%
Alternatively, one could use the on-shell action method by Ba\~nados, Canto and Theisen \cite{Banados:2012ue}.}
\eq{
F(T,\,\Omega,\,\Omega_{\textrm{\tiny V}},\,\Omega_{\textrm{\tiny U}}) = -{\cal Q}_{\cal M} - TS - \Omega\,{\cal Q}_{\cal L} - \Omega_{\textrm{\tiny V}}\,{\cal Q}_{\cal V} - \Omega_{\textrm{\tiny U}}\,{\cal Q}_{\cal U}
}{eq:cp127}
The zero mode charges are given by \eqref{eq:Q} and the intensive quantities by the chemical potentials.
\begin{align}
 T^{-1} = \beta &= -\frac{\partial S}{\partial{\cal Q}_{\cal M}}\Big|_{\cal L,V,U} = -\beta_L\,(1+\chemM)\\
 \beta\,\Omega &= -\frac{\partial S}{\partial{\cal Q}_{\cal L}}\Big|_{\cal M,V,U} = -\beta_L\,\chemL\\
 \beta\,\Omega_{\textrm{\tiny V}} &= -\frac{\partial S}{\partial{\cal Q}_{\cal V}}\Big|_{\cal M,L,U} = -\beta_L\,\chemV\\
 \beta\,\Omega_{\textrm{\tiny U}} &= -\frac{\partial S}{\partial{\cal Q}_{\cal U}}\Big|_{\cal M,L,V} = -\beta_L\,\chemU
\end{align}

In order to express free energy in terms of intensive variables we have to invert the holonomy conditions and solve for the charges in terms of chemical potentials. Before doing so, it is instructive to consider free energy expressed in terms of charges in certain limits. 
In the large ${\cal R}$ limit (weak contribution from spin-3 charges) we recover the spin-2 result
\eq{
F_{\textrm{\tiny weak}} = -\frac{\cal M}{2} + {\cal O}({\cal P}/\sqrt{\cal R}) + {\cal O}(1/{\cal R})\,.
}{eq:cp129}
In the ${\cal R}\to 3$ limit (strong contribution from spin-3 charges) we obtain
\eq{
F_{\textrm{\tiny strong}} = -\frac{\cal M}{6} + {\cal O}({\cal R}-3)^2\,.
}{eq:cp130}
Thus, we have a universal ratio
\eq{
\frac{F_{\textrm{\tiny weak}}}{F_{\textrm{\tiny strong}}} = 3\,.
}{eq:cp131}
The results \eqref{eq:cp129}-\eqref{eq:cp131} are valid on all branches and show that the free energy approaches the correct spin-2 value.

Performing the Legendre transformation \eqref{eq:cp127} with the entropy \eqref{eq:S2} yields
\eq{
F = -{\cal Q}_{\cal M} + T\beta_L\,\chemV {\cal Q}_{\cal V} = - \frac{\cal M}{2} - 4\Omega_{\textrm{\tiny V}}{\cal V}  \,.
}{eq:F}
In order to obtain free energy as function of intensive variable we have to solve the non-linear holonomy conditions \eqref{eq:cp106}, \eqref{eq:cp107} for the charges in terms of the chemical potentials. Solving \eqref{eq:cp106} for ${\cal V}$ allows us to express free energy in terms of the mass ${\cal M}$ and of chemical potentials.
\eq{
F = - \frac{\cal M}{2} + \frac{{\cal M} \Omega \Omega_{\textrm{\tiny V}}}{6 \Omega_{\textrm{\tiny U}}} + \frac{2{\cal M}^2 \Omega_{\textrm{\tiny U}} \Omega_{\textrm{\tiny V}}}{9 \Omega} - \frac{2\pi^2T^2\Omega_{\textrm{\tiny V}}}{3 \Omega \Omega_{\textrm{\tiny U}}}
}{eq:F1}
Plugging the solution for the spin-3 charge $\cal V$ in terms of the mass $\cal M$ into the other holonomy condition \eqref{eq:cp107} establishes a quartic equation for the mass $\cal M$, which leads to four branches of solutions for free energy. The discriminant of that equation is positive, provided the spin-3 chemical potential obeys the bound
\eq{
\Omega_{\textrm{\tiny U}}^2 < \frac{9(2\sqrt{3}-3)}{64}\,\frac{\Omega^4}{4\pi^2 T^2} \approx 0.065 \,\frac{\Omega^4}{4\pi^2 T^2} \,.
}{eq:cp132}
Another way to read the inequality \eqref{eq:cp132} is that it provides an upper bound on the temperature for given spin-3 chemical potential $\Omega_{\textrm{\tiny U}}$. The maximal temperature is given by
\eq{
T_{\textrm{\tiny max}} = \frac{3\sqrt{2\sqrt{3}-3}}{8}\,\frac{\Omega^2}{2\pi |\Omega_{\textrm{\tiny U}}|}\,.
}{eq:cp136}

\newcommand{\stwo}{{\frak t}}
\newcommand{\sthree}{{\frak v}}

In the limit of small $\Omega_{\textrm{\tiny U}}$ it turns out that only one of the branches has finite free energy. This is the branch that continuously connects with spin-2 results, on which free energy yields
\eq{
F = -\frac{2\pi^2 T^2}{\Omega^2}\,\Big(1 - \frac{32\pi^2 T^2 \Omega_{\textrm{\tiny V}} \Omega_{\textrm{\tiny U}}}{
 3 \Omega^3} + \frac{80 \pi^2 T^2 \Omega_{\textrm{\tiny U}}^2}{3 \Omega^4} + {\cal O}(\Omega_{\textrm{\tiny U}}^3)\Big)\,.
}{eq:F2}
The term before the parentheses reproduces the spin-2 result for free energy. The term in the parentheses depends only on two linear combinations of the chemical potentials [on $\stwo$  and $\sthree$ introduced in \eqref{eq:F4} below].
As in the spin-2 case \cite{Bagchi:2013lma} there will be a phase transition between flat space cosmologies and hot flat space at some critical temperature. 

A novel feature of the spin-3 case is that there are additional phase transitions between the various flat space cosmology branches. To see this we consider the difference between the free energies of two branches.
\eq{
\Delta F_{12} = \frac{2\Omega_{\textrm{\tiny U}}\Omega_{\textrm{\tiny V}}}{9\Omega}\,\big({\cal M}_1 - {\cal M}_2\big)\,\Big({\cal M}_1 + {\cal M}_2 +\frac{3 \Omega (\Omega \Omega_{\textrm{\tiny V}}-3\Omega_{\textrm{\tiny U}})}{4 \Omega_{\textrm{\tiny U}}^2 \Omega_{\textrm{\tiny V}}}\Big)
}{eq:F3}
There are two zeros in the difference \eqref{eq:F3}, an obvious one when the masses of the two branches coincide, ${\cal M}_1={\cal M}_2$, and a non-obvious one when the expression in the last parentheses in \eqref{eq:F3} vanishes. We focus in the following on the difference between the branch that continuously connects to spin-2 results (branch 1) and the other branch that ceases to exist if the bound \eqref{eq:cp132} is violated (branch 2). The other two branches are then branch 3 and 4; they will play only minor roles. 

To reduce clutter we assume from now on that temperature and the chemical potentials are non-negative. Moreover, we introduce
dimensionless combinations of chemical potentials
\eq{
\stwo = 2\pi T\,\frac{\Omega_{\textrm{\tiny U}}}{\Omega^2} \qquad \sthree = \Omega_{\textrm{\tiny V}}\,\frac{\Omega}{\Omega_{\textrm{\tiny U}}}\,.
}{eq:F4}
The quantity $\stwo$ is a dimensionless temperature, while $\sthree$ is essentially a ratio of odd over even spin-3 chemical potential.
Expressing the difference of free energies \eqref{eq:F3} between branches 1 and 2 as function of these two combinations, up to a non-negative overall constant, yields
\eq{
\Delta F_{12} \propto 15\sthree - 18 - \sthree\, \sqrt{64 \stwo^2 + 9 + \frac{8 \stwo (64 \stwo^2 + 27)}{N(\stwo)} +  8 \stwo N(\stwo)}\,.
}{eq:cp137}
with 
\eq{
N(\stwo) = \big(512 \stwo^3 + 648 \stwo +  9 \sqrt{4096 \stwo^4 + 3456 \stwo^2 - 243  }\big)^{1/3}\,.
}{eq:cp138}
The positive real zero of the term under the square-root in \eqref{eq:cp138} corresponds precisely to the critical temperature \eqref{eq:cp136}.
For each value of dimensionless temperature $\stwo$ there is a simple zero in $\Delta F_{12}$ since it depends linearly on $\sthree$. We call the corresponding value of $\sthree$ `critical' and denote it by subscript `c'. For vanishing temperature we find from equating \eqref{eq:cp137} to zero
\eq{
\sthree_c|_{\stwo=0} = \frac32
}{eq:cp139}
while at the critical temperature \eqref{eq:cp136} we find similarly 
\eq{
\sthree_c|_{\stwo=\stwo_c=\tfrac38 \sqrt{2 \sqrt{3} - 3}} = 2\,.
}{eq:cp140}
The corresponding free energy differences near these temperatures read, respectively
\begin{align}
 \Delta F_{12} &\propto 12 \sthree - 18 - 12 \stwo\sthree + \tfrac83 \stwo^2\sthree + {\cal O}(\stwo^3)\\
 \Delta F_{12} &\propto 9 \sthree - 18 - 8\sqrt{1+\tfrac{2}{\sqrt{3}}}\,(\stwo-\stwo_c) \sthree - \tfrac{16}{27}\, (\stwo-\stwo_c)^2\,\sthree + {\cal O}(\stwo-\stwo_c)^3\,.
\end{align}
We arrive therefore at the following picture, depending on the value of the parameter $\sthree$:\footnote{%
Positivity of entropy imposes additional constraints on the existence of branches; we checked that the existence of the first order phase transition between branches 1 and 2 that we describe below is not influenced by such constraints.}
\begin{itemize}
 \item $0<\sthree<\tfrac32$: Branch 1 is thermodynamically unstable for all temperatures.
 \item $\sthree=\tfrac32$: Branch 1 degenerates with branch 2 at vanishing temperature and is thermodynamically unstable for all positive temperatures.
 \item $\tfrac32<\sthree<2$: Branch 1 degenerates with branch 2 at some positive temperature. Below that temperature branch 1 is thermodynamically unstable. At that temperature there is a phase transition from branch 2 to branch 1. Above that temperature branch 1 is stable (modulo the phase transition to hot flat space \cite{Bagchi:2013lma}).
 \item $\sthree=2$: Branch 1 degenerates with branch 2 at the maximal temperature \eqref{eq:cp136} and is thermodynamically stable for all temperatures (again modulo the phase transition to hot flat space).
 \item $\sthree>2$: Branch 1 is thermodynamically stable for all temperatures (with the same caveat as above).
\end{itemize}

\enlargethispage{0.5truecm}

\begin{figure}
\begin{center}
\includegraphics[width=0.32\linewidth]{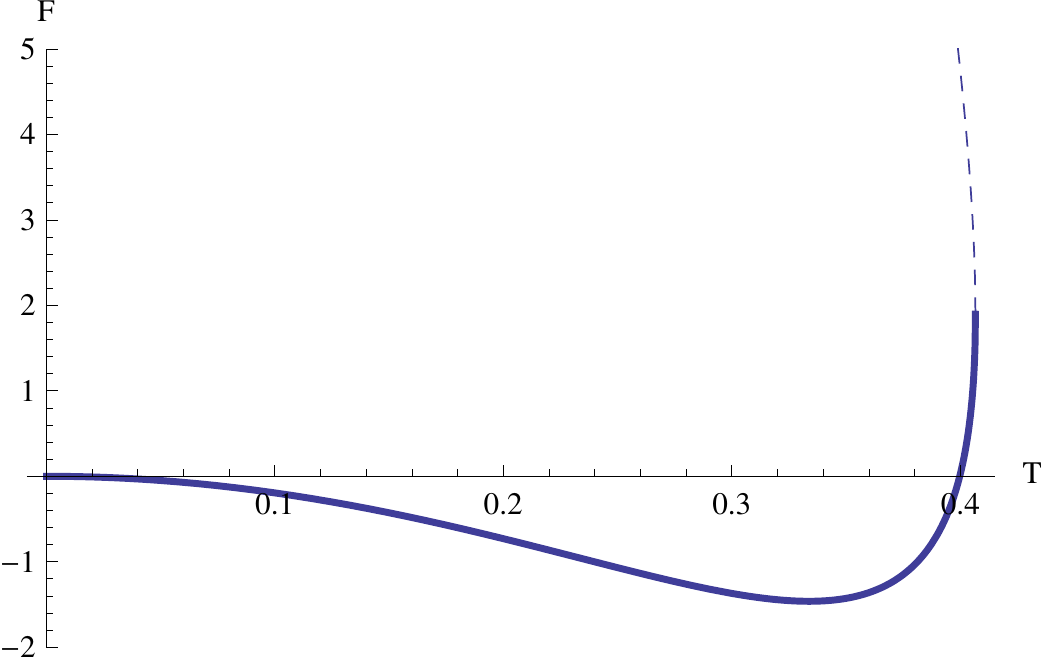}
\includegraphics[width=0.32\linewidth]{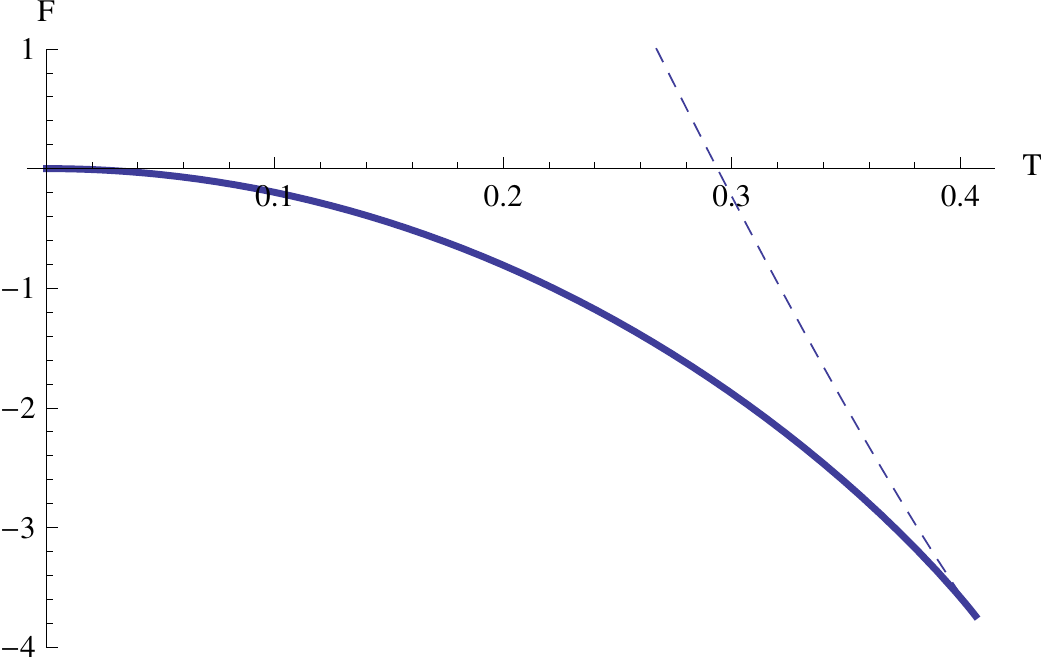}
\includegraphics[width=0.32\linewidth]{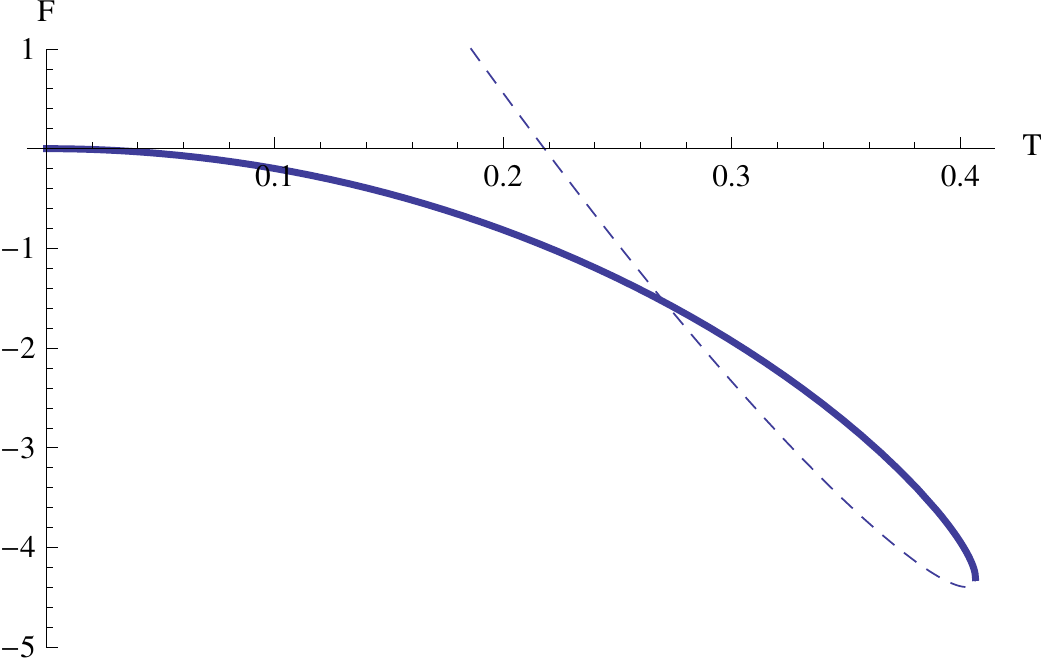}
\includegraphics[width=0.32\linewidth]{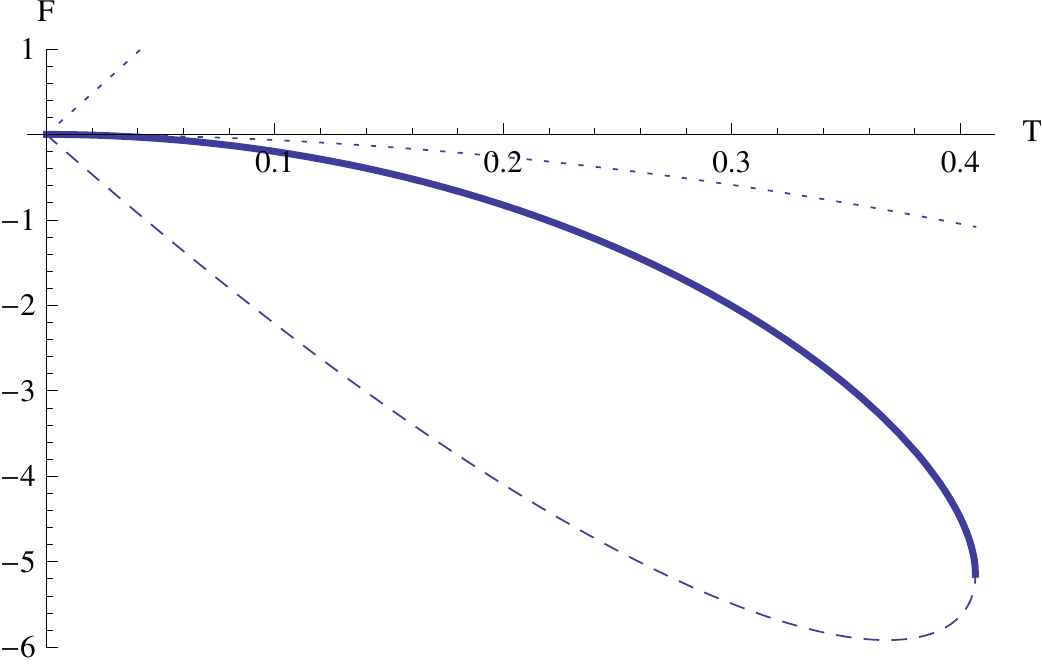}
\includegraphics[width=0.32\linewidth]{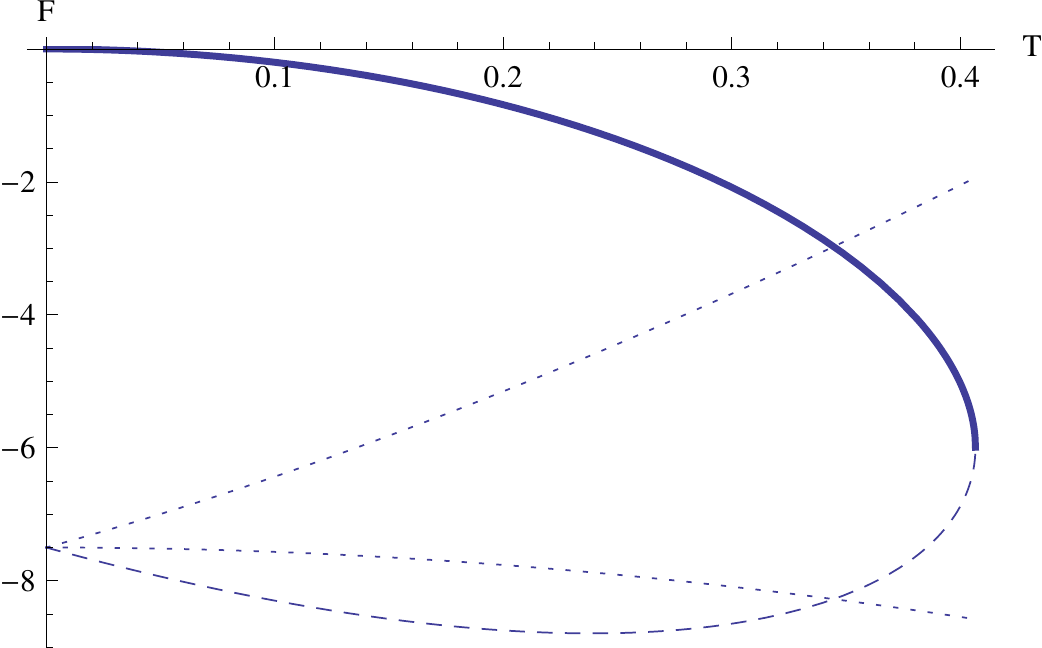}
\includegraphics[width=0.32\linewidth]{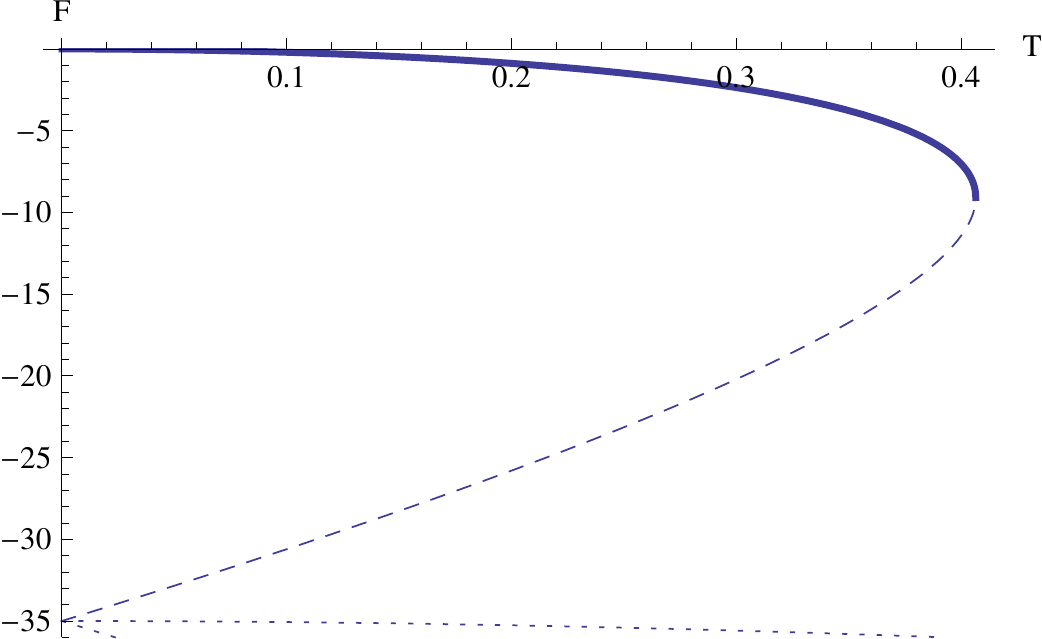}
 \caption[Plots of free energy as function of temperature.]{Plots of free energy as function of temperature. In all plots 
$\Omega=1$, $\Omega_{\textrm{\tiny U}}=0.1$.
Upper Left: $\Omega_{\textrm{\tiny V}}=0.4$. Upper Middle: $\Omega_{\textrm{\tiny V}}=0.2$. Upper Right: $\Omega_{\textrm{\tiny V}}=0.18$. 
Lower Left: $\Omega_{\textrm{\tiny V}}=0.15$. Lower Middle: $\Omega_{\textrm{\tiny V}}=0.12$. Lower Right: $\Omega_{\textrm{\tiny V}}=0.01$. 
The branch with smooth spin-2 limit is displayed as thick line, the second branch as dashed line, the other two branches as dotted lines (in the upper plots these lines are at positive $F$). For a movie of these plots see \href{http://quark.itp.tuwien.ac.at/\~grumil/mp3/free\_energy\_fs3.avi}{{\tt http://quark.itp.tuwien.ac.at/$\sim$grumil/mp3/free\_energy\_fs3.avi}}.}
\label{fig:1}
\end{center}
\end{figure}

To illustrate the results above we show an example in figure \ref{fig:1}. In all six graphs the thick line depicts free energy for branch 1 and the dashed line for branch 2 (the other two branches are not essential for this discussion; if visible they are plotted as dotted lines). 
The three upper plots show explicitly the phase transition between branches 1 and 2, depending on the choice of $\sthree$. The three lower plots show that there are further phase transitions involving the branches 3 and 4, if branch 1 is unstable for all values of temperature. In addition to all these new phase transitions there is the `usual' phase transition to hot flat space \cite{Bagchi:2013lma}, which in the present case can be of zeroth, first or second order. Since there are several phase transitions possible there exist also multi-critical points where three or four phases co-exist. 

The most striking difference between the AdS results by David, Ferlaino and Kumar \cite{David:2012iu} and our flat space results is that we observe the possibility of first order phase transitions between various branches (see the right upper and middle lower plot in figure \ref{fig:1}). By contrast, in AdS the only phase transitions (other than Hawking--Page like) arise because two of the branches end, at which point the free energy jumps (we also recover these zeroth order phase transitions in flat space, see e.g.~the left lower plot in figure \ref{fig:1}). 

\subsection{Remarks on flat space singularity resolutions}\label{se:4.4}

String theory is believed to resolve (some of) the singularities that arise in classical gravity, see e.g.~\cite{Natsuume:2001ba} and references therein. If this is true and if higher spin gravity can be thought of as emerging from string theory in the tensionsless limit, then it is suggestive that also higher spin gravity could resolve (some of) the singularities that arise in Einstein gravity. Regardless of how plausible this line of reasoning appears, it is certainly of interest to investigate the issue of singularity resolution in (three-dimensional) higher spin gravity.

Indeed, Castro et al.~discovered corresponding singularity resolutions for black holes \cite{Castro:2011fm} and conical surpluses \cite{Castro:2011iw} in three-dimensional AdS higher spin gravity. More recently, Krishnan et al.~considered the singularity resolutions of the Milne universe \cite{Krishnan:2013tza} and the null orbifold \cite{Kiran:2014kca} in three-dimensional flat space higher spin gravity.  We discuss now some of their findings from the perspective developed in the present work, starting with the second example, the null orbifold singularity resolution.

In our conventions the null orbifold is a configuration with ${\cal M}={\cal N}={\cal V}={\cal Z}=\chemM=\chemL=\chemV=\chemU=0$, i.e.
\eq{
a_u = M_+\qquad a_\varphi = L_+\,.
}{eq:cp42}
This configuration leads to the null orbifold line-element
\eq{
\extd s^2 = -2\extd r\extd u +r^2\,\extd\varphi^2
}{eq:cp43}
with vanishing spin-3 field. The null orbifold exhibits a singularity at $r=0$, see e.g.~\cite{Horowitz:1990ap,FigueroaO'Farrill:2001nx,Liu:2002ft,Simon:2002ma}.

One of the claims of \cite{Kiran:2014kca} is that there is a spin-3 gauge transformation that resolves this singularity. In the language of the present work, this resolution involves the following connection
\eq{
a_u = M_+\qquad a_\varphi = L_+ + \tfrac92\,p\, V_0 
}{eq:cp44}
which leads to the line-element
\eq{
g_{\mu\nu}\,\extd x^\mu\extd x^\nu = -2\extd r\extd u + \big(r^2+27p^2\big)\,\extd\varphi^2
}{eq:cp45}
and spin-3 field
\eq{
\Phi_{\mu\nu\la}\,\extd x^\mu\extd x^\nu\extd x^\lambda = 3p\,\big(\extd r\extd u\extd\varphi + (r^2-9p^2)\,\extd\varphi^3\big)\,.
}{eq:cp46}
Up to a different choice of coordinates and overall normalization of the spin-3 field, the results \eqref{eq:cp45} and \eqref{eq:cp46} coincide precisely with Eq.~(3.16) in \cite{Kiran:2014kca}.

Comparing the original null orbifold configuration \eqref{eq:cp42} with the resolved one \eqref{eq:cp44} we see that the difference is in the $a_\varphi$ component, not the $a_u$ component. Therefore, we cannot interpret the additional terms proportional to $p$ as coming from a chemical potential as introduced in section \ref{se:3}. 

We check now whether the (spin-3) transformation that maps \eqref{eq:cp42} to \eqref{eq:cp44} is a small gauge transformation. If $p$ is a state-dependent function then the term proportional to $p$ in \eqref{eq:cp44} leads to a contribution to the canonical currents of the form (see appendix \ref{se:4.3})
\eq{
\delta Q \sim \oint \extd\varphi \,(\chi''-{\cal M}\chi)\delta p
}{eq:cp47}
which is finite and conserved, but not integrable in field space unless $p=p({\cal M})$. However, if $p$ is not a state-dependent parameter but merely some (gauge-)parameter then its field-variation vanishes, $\delta p=0$, and the canonical charges remain unchanged. 

Therefore, we conclude that the spin-3 singularity resolution discussed in \cite{Kiran:2014kca} is based on a small gauge transformation, i.e., it neither changes the canonical charges nor the chemical potentials.
Our results thus support their conclusions.  

The same remarks apply to the spin-3 singularity resolution of the Milne universe \cite{Krishnan:2013tza}, which manifestly uses a gauge transformation of the type discussed in appendix \ref{se:4.3}, with state-independent constant gauge parameters $v_0\neq 0 \neq v_2$.

\subsection{Further applications, developments and generalizations}\label{se:4.5}

Above we have presented some applications of flat space (spin-3) gravity with chemical potentials. Below we mention several other possible applications and generalizations that we leave for future work.
\begin{itemize}
 \item {\bf Flat space higher spin Cardy formula.} The usual Cardy formula \cite{Cardy:1986ie,Strominger:1997eq} was generalized in (at least) two ways: 1.~by including higher spin fields \cite{Kraus:2011ds,Gaberdiel:2012yb,deBoer:2013gz} and 2.~by taking the flat space limit \cite{Barnich:2012xq,Bagchi:2012xr,Riegler:2014bia,Fareghbal:2014qga}. It seems both natural and interesting to combine these two generalizations and to derive a Cardy-like formula for the entropy of spin-3 flat space cosmologies that (hopefully) matches our result \eqref{eq:S}.
 \item {\bf Flat space family of solutions beyond flat space cosmologies.} It could be rewarding to study in detail solutions of the holonomy conditions \eqref{eq:cp105} for integers $m\neq 2$ and $n\neq 1$. The ensuing family of solutions could play an analogous role for flat space (higher spin) gravity as the SL$(2,\mathbb{Z})$ family of Euclidean saddlepoints in AdS spin-2 gravity. 
 \item {\bf Flat space spin-3 holographic dictionary.} Following our discussion in section \ref{sec:E} it would be interesting to continue the flat space holographic dictionary, in particular by identifying the sources (or non-normalizable modes) for the spin-3 field. To this end we linearize the result \eqref{eq:cp21} in $\chemV$ and $\chemU$.
\eq{
\Phi_{\mu\nu\lambda} \extd x^\mu \extd x^\nu \extd x^\lambda = \bar \Phi_{\mu\nu\lambda} \extd x^\mu \extd x^\nu \extd x^\lambda + \Psi_{\mu\nu\lambda} \extd x^\mu \extd x^\nu \extd x^\lambda + {\cal O}(\chemV^2,\chemU^2,\chemV\chemU)
}{eq:cp67}
For simplicity, we set to zero the spin-2 chemical potentials and charges, as well as the spin-3 charges, and assume that all the spin-3 chemical potentials are constant. The background solution $\bar \Phi_{\mu\nu\lambda} \extd x^\mu \extd x^\nu \extd x^\lambda $ is given by the right hand side of \eqref{eq:cp16}.
With these assumptions, the linear piece in the chemical potentials yields
\eq{
\Psi_{\mu\nu\lambda} \extd x^\mu \extd x^\nu \extd x^\lambda = - 2r^2\, \chemU \, \extd r  \extd u \extd\varphi + \chemV\, \extd r^2 \extd u  \,.
}{eq:cp68} 
By analogy to the discussion after \eqref{eq:cp66}, we conjecture that the two terms in \eqref{eq:cp68} should correspond to the essential pieces in the two towers of non-normalizable solutions to the linearized spin-3 equations of motion. 
 \item {\bf isl$\boldsymbol{(N)}$.} Everything we have done in the present work should generalize straightforwardly to higher spin gravity theories based on an isl$(N)$ connection, with $N>3$. In fact, for the principal embedding we expect that all our conclusions remain essentially unchanged. All flat space results should be obtainable from a suitable \.In\"on\"u--Wigner contraction of corresponding AdS results based on an sl$(N)\,\oplus$\,sl$(N)$ connection. Similar generalizations in AdS were considered in \cite{Chen:2012ba,Ferlaino:2013vga,Beccaria:2013gaa}.
 \item {\bf Non-principal embeddings.} Whenever the corresponding AdS results are known, again we expect that all flat space results should be obtainable from a suitable \.In\"on\"u--Wigner contraction. There could be interesting surprises for non-principal embeddings in the flat space limit, however, as the discussion in \cite{Grumiller:2014lna} shows. 
 \item {\bf Gravitational anomalies.} It is of interest to generalize the result \eqref{eq:cp124} for entropy to theories which are obtained as flat space \.In\"on\"u--Wigner contractions from AdS theories with gravitational anomalies so that $c_L = c - \bar c \neq 0$. In \cite{Riegler:2014bia} such an expression was found, which correctly reproduces \eqref{eq:cp124} (up to a different choice of normalization of $\mathcal{L}$ and $\mathcal{U}$). Moreover, it also gives a prediction for the thermal entropy of flat space cosmology solutions in the presence of gravitational anomalies. This result can be obtained using the methods presented in section \ref{se:4.2} upon replacing the hatted trace with (one quarter of the) trace and the level $k$ with $c_L/24$.
 \item {\bf First order phase transitions in AdS higher spin gravity.} Some of our results resemble corresponding AdS results. For instance, the branch that continuously connects to spin-2 gravity also becomes unstable beyond a critical temperature in AdS \cite{David:2012iu}. Moreover, this temperature agrees quantitatively with our result \eqref{eq:cp136}, upon replacing our ratio $\Omega^2/|\Omega_{\textrm{\tiny U}}|$ by their $\mu^{-1}$. However, the first order phase transitions discovered in section \ref{se:4.new} do not arise in AdS, despite of the fact that the main ingredient we used was to solve the non-linear holonomy conditions \eqref{eq:cp106}, \eqref{eq:cp107} for the charges in terms of chemical potentials, and these holonomy conditions are identical to the ones in AdS higher spin gravity \cite{Bunster:2014mua}. It could be interesting to make a scan through all possibilities in AdS higher spin gravity to see if some novel first order phase transitions can arise, and if not, to understand 
better why AdS and flat space behave so differently in this regard. 
 \item {\bf Holographic entanglement entropy.} Entanglement entropy of Galilean CFTs, the dual field theories that arise in flat space spin-2 gravity, was derived recently \cite{Bagchi:2014iea}. It would be very interesting to generalize the discussion to the spin-3 case (or even higher spins), both on the field theory and the higher spin gravity sides, along the lines of \cite{Ammon:2013hba,deBoer:2013vca} or by suitably contracting the results of \cite{Datta:2014uxa}.
\end{itemize}

\enlargethispage{1truecm}


\section*{Acknowledgments}

We thank Hamid Afshar, Arjun Bagchi, Geoffrey Comp\`ere, Stephane Detournay, Reza Fareghbal, Wout Merbis, Stefan Prohazka, Friedrich Sch{\"o}ller and Joan Simon for discussions.

This work was supported by the START project Y~435-N16 of the Austrian Science Fund (FWF) and the FWF projects I~952-N16, I~1030-N27 and P~27182-N27. During the final stages MG was supported by the FWF project P~27396-N27.

\begin{appendix}
 
\section{Matrix representations of isl$\boldsymbol{(3)}$ generators}\label{app:A}

\subsection[6x6 representation]{$\boldsymbol{6\times 6}$ representation}\label{app:A.1}

In most of our work we use the following matrix representation of isl$(3)$ generators in terms of $6\times 6$ block-diagonal matrices. It is convenient to write them as a $3\times 3$ block tensored by a simple diagonal $2\times 2$ matrix. The block structure is a remnant of the decomposition of the AdS algebra so$(2,2)\sim \textrm{so}(2,1)\oplus \textrm{so}(2,1)$ before the \.In\"on\"u--Wigner contraction. 

Even spin-2 generators:
\eq{
 L_+ = \begin{pmatrix}
        0 & 0 & 0   \\
        1 & 0 & 0   \\
        0 & 1 & 0  
       \end{pmatrix} \otimes \unity_{2\times 2} \qquad 
 L_0 = \begin{pmatrix}
        1 & 0 & 0  \\
        0 & 0 & 0 \\
        0 & 0 & -1 
       \end{pmatrix}  \otimes \unity_{2\times 2}\qquad 
 L_- = \begin{pmatrix}
        0 & -2 & 0  \\
        0 & 0 & -2  \\
        0 & 0 & 0   
       \end{pmatrix} \otimes \unity_{2\times 2}
}{eq:app1}
Even spin-3 generators:
\begin{align}
 & U_2 = \begin{pmatrix}
        0 & 0 & 0 \\
        0 & 0 & 0 \\
        2 & 0 & 0 
       \end{pmatrix} \otimes \unity_{2\times 2} \quad 
 & U_1 = \begin{pmatrix}
        0 & 0 & 0 \\
        1 & 0 & 0 \\
        0 & -1 & 0 
       \end{pmatrix} \otimes \unity_{2\times 2} \qquad 
 & U_0 = \begin{pmatrix}
        \tfrac23 & 0 & 0  \\
        0 & -\tfrac43 & 0  \\
        0 & 0 & \tfrac23  
       \end{pmatrix} \otimes \unity_{2\times 2} \nonumber \\
 && U_{-1} = \begin{pmatrix}
        0 & -2 & 0 \\
        0 & 0 & 2  \\
        0 & 0 & 0 
       \end{pmatrix} \otimes \unity_{2\times 2}\qquad 
 & U_{-2} = \begin{pmatrix}
        0 & 0 & 8  \\
        0 & 0 & 0  \\
        0 & 0 & 0 
       \end{pmatrix} \otimes \unity_{2\times 2}
 \label{eq:app3}
\end{align}
All odd generators can be written as a product of corresponding even generators times a $\gamma^\ast$-matrix,
\eq{
M_n = \epsilon\,L_n \times \gamma^\ast \qquad V_n = \epsilon\,U_n \times \gamma^\ast
}{eq:app5}
with $\epsilon$ a Grassmann-parameter, $\epsilon^2=0$, and
\eq{
\gamma^\ast = \begin{pmatrix}
               \unity_{3\times 3} & \mathbb{O}_{3\times 3} \\
               \mathbb{O}_{3\times 3} & -\unity_{3\times 3}  
              \end{pmatrix}\,.
}{eq:app6}
Equivalently, one can replace in the formulas \eqref{eq:app1}, \eqref{eq:app3} everywhere the factor $\unity_{2\times 2}$ by the diagonal Pauli matrix $\sigma_3 = \textrm{diag}(1,\,-1)$ times the Grassmann parameter $\epsilon$ in order to obtain the odd generators from the corresponding even ones.

\subsection[8+1 representation]{$\boldsymbol{8+1}$ representation}\label{app:A.2}

For deriving entropy and holonomy conditions we use the following matrix representation of isl$(3)$ generators in terms of $8+1$-dimensional matrices with a ``tensor''- and a ``vector''-block. Generic generators $G$ are written in the form
\eq{
G = \begin{pmatrix}
     \textrm{ad}_{8\times 8} & \textrm{odd}_{8\times 1} \\
     \mathbb{O}_{1\times 8} & 0
    \end{pmatrix}
}{eq:ad1}
where $\textrm{ad}_{8\times 8}$ is an $8\times 8$ matrix that is an element of sl$(3)$ in the adjoint representation and $\textrm{odd}_{8\times 1}$ is an $8\times 1$ column vector. The even generators $L_n$ and $U_n$ have $\textrm{ad}\neq \mathbb{O}$, $\textrm{odd}=\mathbb{O}$; the odd generators $M_n$ and $V_n$ have $\textrm{ad}=\mathbb{O}$, $\textrm{odd}\neq \mathbb{O}$. In fact, we can (and will) use the odd generators as unit basis vectors, 
\eq{\textrm{odd}_{M_n}=E_{n+2}\qquad \textrm{odd}_{V_n}=E_{n+6}
}{eq:ad3}
with
\eq{
E_i=(\underbrace{0,\dots,0}_{i-1},1,\underbrace{0,\dots,0}_{8-i})^T\qquad i=1..8\,.
}{eq:ad2}
The $\textrm{ad}$-parts of the even generators compatible with the algebra \eqref{eq:FSHSG14} are then given by the following $8\times8$ matrices.
{\footnotesize
\begin{align*}
 \textrm{ad}_{L_{-1}} &= -\begin{pmatrix}
            0 & 1 & 0 & 0 & 0 & 0 & 0 & 0 \\
            0 & 0 & 2 & 0 & 0 & 0 & 0 & 0 \\
            0 & 0 & 0 & 0 & 0 & 0 & 0 & 0 \\
            0 & 0 & 0 & 0 & 1 & 0 & 0 & 0 \\
            0 & 0 & 0 & 0 & 0 & 2 & 0 & 0 \\
            0 & 0 & 0 & 0 & 0 & 0 & 3 & 0 \\
            0 & 0 & 0 & 0 & 0 & 0 & 0 & 4 \\
            0 & 0 & 0 & 0 & 0 & 0 & 0 & 0 
           \end{pmatrix} &
 \textrm{ad}_{L_0} &= \begin{pmatrix}
            1 & 0 & 0 & 0 & 0 & 0 & 0 & 0 \\
            0 & 0 & 0 & 0 & 0 & 0 & 0 & 0 \\
            0 & 0 & -1 & 0 & 0 & 0 & 0 & 0 \\
            0 & 0 & 0 & 2 & 0 & 0 & 0 & 0 \\
            0 & 0 & 0 & 0 & 1 & 0 & 0 & 0 \\
            0 & 0 & 0 & 0 & 0 & 0 & 0 & 0 \\
            0 & 0 & 0 & 0 & 0 & 0 & -1 & 0 \\
            0 & 0 & 0 & 0 & 0 & 0 & 0 & -2 
         \end{pmatrix} &
 \textrm{ad}_{L_1} &= \begin{pmatrix}
            0 & 0 & 0 & 0 & 0 & 0 & 0 & 0 \\
            2 & 0 & 0 & 0 & 0 & 0 & 0 & 0 \\
            0 & 1 & 0 & 0 & 0 & 0 & 0 & 0 \\
            0 & 0 & 0 & 0 & 0 & 0 & 0 & 0 \\
            0 & 0 & 0 & 4 & 0 & 0 & 0 & 0 \\
            0 & 0 & 0 & 0 & 3 & 0 & 0 & 0 \\
            0 & 0 & 0 & 0 & 0 & 2 & 0 & 0 \\
            0 & 0 & 0 & 0 & 0 & 0 & 1 & 0 
           \end{pmatrix}  \displaybreak[1] \\
 \textrm{ad}_{U_{-2}} &= \begin{pmatrix}
            0 & 0 & 0 & 0 & 0 & 0 & 4 & 0 \\
            0 & 0 & 0 & 0 & 0 & 0 & 0 & 16 \\
            0 & 0 & 0 & 0 & 0 & 0 & 0 & 0 \\
            0 & -2 & 0 & 0 & 0 & 0 & 0 & 0 \\
            0 & 0 & -4 & 0 & 0 & 0 & 0 & 0 \\
            0 & 0 & 0 & 0 & 0 & 0 & 0 & 0 \\
            0 & 0 & 0 & 0 & 0 & 0 & 0 & 0 \\
            0 & 0 & 0 & 0 & 0 & 0 & 0 & 0 
         \end{pmatrix} &
 \textrm{ad}_{U_{-1}} &= \begin{pmatrix}
            0 & 0 & 0 & 0 & 0 & -2 & 0 & 0 \\
            0 & 0 & 0 & 0 & 0 & 0 & -2 & 0 \\
            0 & 0 & 0 & 0 & 0 & 0 & 0 & 4 \\
            1 & 0 & 0 & 0 & 0 & 0 & 0 & 0 \\
            0 & -1 & 0 & 0 & 0 & 0 & 0 & 0 \\
            0 & 0 & 1 & 0 & 0 & 0 & 0 & 0 \\
            0 & 0 & 0 & 0 & 0 & 0 & 0 & 0 \\
            0 & 0 & 0 & 0 & 0 & 0 & 0 & 0 
           \end{pmatrix} &
 \textrm{ad}_{U_0} &= \begin{pmatrix}
            0 & 0 & 0 & 0 & 2 & 0 & 0 & 0 \\
            0 & 0 & 0 & 0 & 0 & 0 & 0 & 0 \\
            0 & 0 & 0 & 0 & 0 & 0 & -2 & 0 \\
            0 & 0 & 0 & 0 & 0 & 0 & 0 & 0 \\
            2 & 0 & 0 & 0 & 0 & 0 & 0 & 0 \\
            0 & 0 & 0 & 0 & 0 & 0 & 0 & 0 \\
            0 & 0 & -2 & 0 & 0 & 0 & 0 & 0 \\
            0 & 0 & 0 & 0 & 0 & 0 & 0 & 0 
         \end{pmatrix}  \displaybreak[1] \\
&&
 \textrm{ad}_{U_1} &= \begin{pmatrix}
            0 & 0 & 0 & -4 & 0 & 0 & 0 & 0 \\
            0 & 0 & 0 & 0 & 2 & 0 & 0 & 0 \\
            0 & 0 & 0 & 0 & 0 & 2 & 0 & 0 \\
            0 & 0 & 0 & 0 & 0 & 0 & 0 & 0 \\
            0 & 0 & 0 & 0 & 0 & 0 & 0 & 0 \\
            3 & 0 & 0 & 0 & 0 & 0 & 0 & 0 \\
            0 & 1 & 0 & 0 & 0 & 0 & 0 & 0 \\
            0 & 0 & -1 & 0 & 0 & 0 & 0 & 0 
           \end{pmatrix} &
 \textrm{ad}_{U_2} &= \begin{pmatrix}
            0 & 0 & 0 & 0 & 0 & 0 & 0 & 0 \\
            0 & 0 & 0 & -16 & 0 & 0 & 0 & 0 \\
            0 & 0 & 0 & 0 & -4 & 0 & 0 & 0 \\
            0 & 0 & 0 & 0 & 0 & 0 & 0 & 0 \\
            0 & 0 & 0 & 0 & 0 & 0 & 0 & 0 \\
            0 & 0 & 0 & 0 & 0 & 0 & 0 & 0 \\
            4 & 0 & 0 & 0 & 0 & 0 & 0 & 0 \\
            0 & 2 & 0 & 0 & 0 & 0 & 0 & 0 
         \end{pmatrix} 
\end{align*}
}

\section{Non-constant contributions to spin-2 and spin-3 fields}\label{app:B}

In this appendix we collect contributions that vanish identically for zero mode solutions with constant chemical potentials,  ${\cal M}^\prime={\cal N}^\prime=\chemM^\prime=\chemL^\prime=\chemV^\prime=\chemU^\prime=0$. We start with expressions for the metric appearing in \eqref{eq:cp20aa}.
\begin{subequations}
 \label{eq:app42}
\begin{align}
g_{uu}^{(r)} &=  \tfrac{16}{3}\,{\cal M}\, \big(\chemU^\prime \chemV - \chemU \chemV^\prime\big) - \tfrac83{\cal N}\chemU^\prime \chemU  - \tfrac83{\cal N}^\prime \chemU^2  \nonumber \\
&\quad  +2\, \big(\chemL^\prime (1+\chemM) - \chemL \chemM^\prime\big)  - \tfrac43\,\big( \chemU''' \chemV  - \chemU \chemV'''\big) + 2\,\big(\chemU'' \chemV^\prime -\chemU^\prime \chemV'' \big) \\
g_{uu}^{(0^\prime)} &= - \tfrac23{\cal M}'' \chemV^2 - \tfrac43 {\cal N}'' \chemU \chemV  - \tfrac53{\cal M}^\prime \chemV^\prime \chemV + \tfrac43{\cal N}^\prime \big( \chemU \chemV^\prime- \tfrac72 \chemU^\prime \chemV\big) - \tfrac53 {\cal M} \big(2 \chemV\chemV''  - \chemV^{\prime\,2}\big) \nonumber \\ 
&\quad - \tfrac43{\cal N}\big( \chemU \chemV'' \!-\! \tfrac52 \chemU^\prime \chemV^\prime \!+\! 4 \chemU'' \chemV \big) - 2(1\!+\!\chemM)\chemM'' + \chemM^{\prime\,2} 
+ \tfrac23(\chemV  \chemV'''' \!-\! \chemV^\prime \chemV''')  + \tfrac13\chemV''^{\,2}  
 \,.
\end{align}
\end{subequations}

We continue with the spin-3 field.
The four coefficient-functions in $\Phi_{uuu}$ contained in \eqref{eq:cp28a} read explicitly
\begin{subequations}
 \label{eq:cp29a}
\begin{align}
\Phi_{uuu}^{(r^3)} &= - \tfrac43 {\cal M}^\prime \chemU^3 - \tfrac43 {\cal M} \chemU^2 \chemU^\prime - \chemL^2 \chemU^\prime + 2 \chemL \chemL^\prime \chemU    + \tfrac43 \chemU^2 \chemU''' - 2 \chemU \chemU^\prime \chemU''  + \chemU^{\prime\,3} \\
\Phi_{uuu}^{(r^2)} &= \Phi_{uuu}^{(r^2,Q'')} + \Phi_{uuu}^{(r^2,Q^\prime)} + \Phi_{uuu}^{(r^2,Q)} + \Phi_{uuu}^{(r^2,\textrm{rest})}  \\ 
\Phi_{uuu}^{(r)} &= \Phi_{uuu}^{(r,Q^2)} + \Phi_{uuu}^{(r,Q\cdot Q^\prime)} + \Phi_{uuu}^{(r,Q'')} + \Phi_{uuu}^{(r,Q^\prime)} + {\cal M}\Phi_{uuu}^{(r,{\cal M})} +   {\cal N}\Phi_{uuu}^{(r,{\cal N})} \nonumber \\
&\quad + {\cal V}\Phi_{uuu}^{(r,{\cal V})}  +  {\cal Z}\Phi_{uuu}^{(r,{\cal Z})} + \Phi_{uuu}^{(r,\textrm{rest})} \\
\Phi_{uuu}^{(0)} &=   
\Phi_{uuu}^{(Q^2)} + \Phi_{uuu}^{(Q\cdot Q)} + {\cal M}''\Phi_{uuu}^{({\cal M}'')}  +  {\cal N}''\Phi_{uuu}^{({\cal N}'')} +  {\cal M}^\prime\Phi_{uuu}^{({\cal M}^\prime)}  +  {\cal N}^\prime\Phi_{uuu}^{({\cal N}^\prime)} \nonumber \\
&\quad  + {\cal M}\Phi_{uuu}^{({\cal M})} + {\cal N}\Phi_{uuu}^{({\cal N})} + {\cal V}\Phi_{uuu}^{({\cal V})} + {\cal Z}\Phi_{uuu}^{({\cal Z})} +  \Phi_{uuu}^{(\textrm{rest})}  
\end{align}
\end{subequations}
with the quadratic part 
\begin{subequations}
 \label{eq:cp29b}
\begin{align}
  \Phi_{uuu}^{(r^2,Q'')} &= \tfrac23 {\cal M}'' \chemU^2 \chemV   + \tfrac43 {\cal N}'' \chemU^3 \\
  \Phi_{uuu}^{(r^2,Q^\prime)} &= {\cal M}^\prime \chemU \big(\tfrac{11}{3} \chemU \chemV^\prime - 2\chemU^\prime \chemV\big) + \tfrac{10}{3} {\cal N}^\prime \chemU^2 \chemU^\prime  \displaybreak[1] \\  
  \Phi_{uuu}^{(r^2,Q)} &=  {\cal M} \big(\chemU^2 \chemV'' + \tfrac43\chemU \chemU'' \chemV + \tfrac23\chemU \chemU^\prime \chemV^\prime - \tfrac73 \chemU^{\prime\,2} \chemV   \big)    + 4 {\cal N} \big(\chemU^2 \chemU'' - \tfrac13 \chemU \chemU^{\prime\,2}\big)  \displaybreak[1] \\
  \Phi_{uuu}^{(r^2,\textrm{rest})} &= (1+\chemM)\big(\chemL^\prime \chemU^\prime - \chemL\chemU''\big) + 2\chemM^\prime \big(\chemL  \chemU^\prime - \chemL^\prime\chemU\big)  - 2 \chemM''\chemL  \chemU  \nonumber \\
&\quad + \chemL^\prime \big(\chemL^\prime\chemV - \chemL \chemV^\prime\big) + \tfrac13 \chemL^2 \chemV''  - \tfrac23 \chemU^2 \chemV''''  - \tfrac43 \chemU \chemU''' \chemV^\prime + \tfrac23 \chemU \chemU'' \chemV''  \nonumber \\
&\quad  + \tfrac23 \chemU \chemU^\prime \chemV^{\prime\,3}  - \chemU^{\prime\,2} \chemV'' - \chemU''^{\,2} \chemV + \chemU'' \chemU^\prime \chemV^\prime  + \tfrac43 \chemU''' \chemU^\prime \chemV  
\end{align}
\end{subequations}
the linear part 
\begin{subequations}
 \label{eq:cp29c}
\begin{align}
\Phi_{uuu}^{(r,Q^2)} &=  \tfrac{16}{9} {\cal M}^2\chemV \big( \chemU \chemV^\prime - \chemU^\prime \chemV\big) + 
  \tfrac{20}{9} {\cal M} {\cal N} \chemU^2 \chemV^\prime - \tfrac83 {\cal M} {\cal N} \chemU \chemU^\prime \chemV - \tfrac89 {\cal N}^2 \chemU^2 \chemU^\prime \\
\Phi_{uuu}^{(r,Q\cdot Q^\prime)} &=  - \tfrac43 {\cal M}^\prime {\cal N} \chemU^2 \chemV + \tfrac89 {\cal M} {\cal N}^\prime \chemU^2 \chemV - \tfrac89 {\cal N} {\cal N}^\prime \chemU^3   \displaybreak[1] \\
\Phi_{uuu}^{(r,Q'')} &= \tfrac23 {\cal M}'' \chemV \big(\chemU^\prime \chemV - \chemU \chemV^\prime\big) + \tfrac43 {\cal N}'' \chemU \big(\chemU^\prime \chemV - \chemU \chemV^\prime\big) \\
\Phi_{uuu}^{(r,Q^\prime)} &= \tfrac13 {\cal M}^\prime \big((1\!+\!\chemM)^2 \chemU - (1\!+\!\chemM) \chemL\chemV   + 2\chemV ( \chemU \chemV'' \!-\! \chemU'' \chemV) + 8 \chemV^\prime ( \chemU^\prime \chemV \!-\! \chemU \chemV^\prime) \big)  \nonumber \\
&\quad  - \tfrac23{\cal N}^\prime \big( (1+\chemM)\chemL \chemU  - 7\chemU^{\prime\,2} \chemV + 2\chemU \chemU'' \chemV  - \tfrac23 \chemU^2 \chemV'' + 6 \chemU \chemU^\prime \chemV^\prime\big)  \displaybreak[1] \\
\Phi_{uuu}^{(r,{\cal M})} &= \tfrac43 (1+\chemM)^2 \chemU^\prime - 2  (1+\chemM) \big(\chemM^\prime \chemU + \tfrac23 (\chemL  \chemV^\prime  - \chemL^\prime \chemV) \big) + \tfrac23  \chemL \chemM^\prime \chemV \nonumber \\
&\quad   + \tfrac{28}{9} \chemU^\prime \chemV \chemV'' - \tfrac{16}{9} \chemU \chemV'' \chemV^\prime - \tfrac43  \chemU'' \chemV \chemV^\prime  - \tfrac49  \chemU \chemV \chemV'''   + \tfrac49 \chemU''' \chemV^2    \\
\Phi_{uuu}^{(r,{\cal N})} &= \tfrac23 (1+\chemM)\big(\chemL^\prime\chemU - \chemL\chemU^\prime\big)  - \tfrac23 \chemL \chemM^\prime \chemU   + 2 \chemL \chemL^\prime \chemV  - \chemL^2 \chemV^\prime + \tfrac89 \chemU \chemU''' \chemV  \nonumber \\
&\quad - \tfrac{14}{3} \chemU \chemU'' \chemV^\prime  + 2 \chemU'' \chemU^\prime \chemV
+ \tfrac49 \chemU^2 \chemV''' + \tfrac53 \chemU^{\prime\,2} \chemV^\prime - \tfrac29 \chemU \chemU^\prime \chemV''  \displaybreak[1] \\
\Phi_{uuu}^{(r,{\cal V})} &=  8  (1+\chemM) \chemU \chemV^\prime - 8  \chemL \chemV \chemV^\prime - 16  \chemM^\prime \chemU \chemV + 16 \chemL^\prime \chemV^2  \\
\Phi_{uuu}^{(r,{\cal Z})} &= 16 (1+\chemM) \chemU \chemU^\prime - 16\chemL \chemV \chemU^\prime   - 32 \chemM^\prime \chemU^2 + 32 \chemL^\prime \chemU \chemV  \displaybreak[1] \\
\Phi_{uuu}^{(r,\textrm{rest})} &=  - \tfrac13(1+\chemM)^2 \chemU''' + (1+\chemM) \big(\chemM^\prime \chemU'' - \chemM'' \chemU^\prime\big) + \tfrac13(1+\chemM)\big( \chemL  \chemV''' - \chemL^\prime \chemV''\big) \nonumber \\
&\quad - \chemM^{\prime\,2} \chemU^\prime + 2 \chemM'' \chemM^\prime \chemU - 2 \chemL^\prime \chemM'' \chemV - \tfrac23 \chemL \chemM^\prime \chemV''  + \chemL \chemM'' \chemV^\prime  + \chemL^\prime \chemM^\prime \chemV^\prime \nonumber \\
&\quad + \tfrac23 \chemU'' \chemV \chemV''' - \tfrac23\chemU^\prime \chemV \chemV''''  - \tfrac49 \chemU''' \chemV \chemV'' - \tfrac29\chemU \chemV''' \chemV'' + \tfrac13\chemU^\prime \chemV''^{\,2}   \nonumber \\
&\quad  - \tfrac13\chemU^\prime \chemV''' \chemV^\prime + \tfrac23 \chemU \chemV'''' \chemV^\prime  - \tfrac13\chemU'' \chemV'' \chemV^\prime  + \tfrac13\chemU''' \chemV^{\prime\,2}  
\end{align}
\end{subequations}
and the constant part
\begin{subequations}
 \label{eq:cp29d}
\begin{align}
%
%
 \Phi_{uuu}^{(Q^2)} &= 
\tfrac23{\cal M}^2 \big(\tfrac53 \chemV^2 \chemV'' \!-  \chemV \chemV^{\prime\,2}\big) - \tfrac49 {\cal N}^2 \big(
 \chemU^2 \chemV'' \!- \tfrac52 \chemU \chemU^\prime \chemV^\prime  
\!- 8 \chemU \chemU'' \chemV\! + \tfrac{25}{4} \chemU^{\prime\,2} \chemV \big) 
\\
\Phi_{uuu}^{(Q\cdot Q)} &= \tfrac49 {\cal M} {\cal N}  \big( 
6 \chemU \chemV \chemV'' - \chemU \chemV^{\prime\,2}  - 5 \chemU^\prime \chemV \chemV^\prime + 4 \chemU'' \chemV^2 \big) 
&  
\\
 \Phi_{uuu}^{({\cal M}'')} &= \tfrac29 {\cal M} \chemV^3 + \tfrac49 {\cal N} \chemU \chemV^2 -\tfrac16(1+\chemM)^2\chemV - \tfrac29 \chemV^2 \chemV'' + \tfrac16 \chemV \chemV^{\prime\,2} \\ 
\Phi_{uuu}^{({\cal N}'')} &= \tfrac49 {\cal M} \chemU \chemV^2 + \tfrac89 {\cal N} \chemU^2 \chemV - \tfrac13 (1+\chemM)^2 \chemU  - \tfrac49 \chemU \chemV \chemV'' + \tfrac13\chemU \chemV^{\prime\,2}  \displaybreak[1] \\
\Phi_{uuu}^{({\cal M}^\prime)} &= -\tfrac19 {\cal M}^\prime \chemV^3 - \tfrac49 {\cal N}^\prime \chemU \chemV^2 + \tfrac13 {\cal M} \chemV^2 \chemV^\prime + \tfrac{16}{9} {\cal N} \chemU \chemV \chemV^\prime - \tfrac{10}{9} {\cal N} \chemU^\prime \chemV^2   \nonumber \\
&\quad - \tfrac{7}{12}(1+\chemM)^2 \chemV^\prime + \tfrac13 (1+\chemM) \chemM^\prime \chemV  + \tfrac29 \chemV^2 \chemV'''  - \tfrac89 \chemV \chemV'' \chemV^\prime + \tfrac{7}{12} \chemV^{\prime\,3} \\
\Phi_{uuu}^{({\cal N}^\prime)} &=  -\tfrac49 {\cal N}^\prime \chemU^2 \chemV + \tfrac89 {\cal M} \chemV \big(\tfrac74 \chemU^\prime \chemV \!-\! \chemU \chemV^\prime \big) + \tfrac89 {\cal N}\chemU \big(2 \chemU^\prime \chemV \!+\! \chemU \chemV^\prime \big) - \tfrac76 (1\!+\!\chemM)^2 \chemU^\prime \nonumber \\
&\quad  + \tfrac23 (1+\chemM) \chemM^\prime \chemU   + \tfrac49 \chemU \chemV \chemV''' - \tfrac29 \chemU \chemV^\prime\chemV'' - \tfrac{14}{9} \chemU^\prime \chemV \chemV''  + \tfrac76 \chemU^\prime \chemV^{\prime\,2} \displaybreak[1] \\
 \Phi_{uuu}^{({\cal M})} &= - \tfrac56(1+\chemM)^2\chemV'' + \tfrac43 (1+\chemM)\chemM^\prime  \chemV^\prime - \tfrac43 (1+\chemM)\chemM'' \chemV  - \tfrac13 \chemM^{\prime\,2} \chemV \nonumber \\
&\quad  -  \tfrac29 \chemV^2 \chemV'''' + \tfrac49 \chemV \chemV^\prime \chemV'''   - \tfrac79 \chemV \chemV''^{\,2}    + \tfrac{7}{18} \chemV^{\prime\,2}\chemV''  \\
\Phi_{uuu}^{({\cal N})} &= - \tfrac43 (1+\chemM)^2 \chemU''  + \tfrac53 (1+\chemM) \chemM^\prime \chemU^\prime - \tfrac23 (1+\chemM) \chemM'' \chemU - \tfrac13 (1+\chemM)\chemL \chemV'' \nonumber \\
&\quad - \tfrac23 \chemM^{\prime\,2} \chemU + \chemM^\prime \chemL  \chemV^\prime  - 2  \chemM'' \chemL \chemV   - \tfrac29 \chemU\big(2\chemV \chemV''''  + \chemV^\prime \chemV''' - \chemV''^{\,2}\big) \nonumber \\
&\quad + \tfrac{10}{9} \chemU^\prime \chemV \chemV''' - \tfrac59 \chemU^\prime\chemV^\prime\chemV''  - \tfrac{16}{9} \chemU'' \chemV \chemV'' + \tfrac43 \chemU'' \chemV^{\prime\,2} \displaybreak[1] \\
\Phi_{uuu}^{({\cal V})} &= 
- 2 (1+\chemM) \chemV^{\prime\,2} + 8 \chemM^\prime \chemV \chemV^\prime  - 16 \chemM'' \chemV^2 \\
\Phi_{uuu}^{({\cal Z})} &= 
- \tfrac{16}{3}(1+\chemM) \chemU \chemV''  + 16 \chemM^\prime \chemU \chemV^\prime - 32 \chemM'' \chemU \chemV  
+ \tfrac{16}{3}\chemL \chemV \chemV'' - 4 \chemL \chemV^{\prime\,2} 
\\
 \Phi_{uuu}^{(\textrm{rest})} &=  \tfrac16(1+\chemM)^2\chemV'''' - \tfrac13(1+\chemM)\chemM^\prime \chemV''' + \tfrac13 (1+\chemM)\chemM'' \chemV'' + \tfrac13\chemM^{\prime\,2} \chemV''  - \chemM^\prime \chemM''  \chemV^\prime \nonumber\\
&\quad  + \chemM''{\,^2} \chemV + \tfrac29 \chemV \chemV'' \chemV'''' - \tfrac19\chemV \chemV'''{\,^2} + \tfrac19 \chemV^\prime\chemV'' \chemV''' - \tfrac16 \chemV^{\prime\,^2}\chemV'''' - \tfrac{1}{27}\chemV''^{\,3}  \,.
\end{align}
\end{subequations}

The remaining non-constant contributions appearing in \eqref{eq:cp28} are given by
\begin{subequations}
 \label{eq:cp29}
\begin{align}
\Phi_{ruu}^{(r)} &= - (1+\chemM) \chemU^\prime + 2 \chemM^\prime \chemU + \chemL \chemV^\prime - 2 \chemL^\prime \chemV \\
\Phi_{ruu}^{(0)} &= \tfrac13(1+\chemM)\chemV'' - \chemM^\prime\chemV^\prime + 2\chemM''\chemV   
\end{align}
\end{subequations}
and
\begin{subequations}
\label{eq:cp29too}
\begin{align}
\Phi_{uu\varphi}^{(r^3)} &= -2\big(\chemL \chemU^\prime - \chemL^\prime \chemU\big) \\
\Phi_{uu\varphi}^{(r^2)} &=   - (1+\chemM) \chemU'' + 2 \chemM^\prime \chemU^\prime - 2 \chemM'' \chemU + \tfrac23 \chemL \chemV'' - \chemL^\prime \chemV^\prime \\ 
\Phi_{uu\varphi}^{(r)} &= - \tfrac43{\cal M}(1+\chemM)\chemV^\prime - \tfrac13{\cal M}^\prime(1+\chemM)\chemV + \tfrac23{\cal M}\chemM^\prime\chemV - 2 {\cal N} (\chemL \chemV^\prime - \chemL^\prime \chemV) \nonumber  \\ 
& - \tfrac23\big( {\cal N} (1+\chemM) \chemU\big)^\prime - 8{\cal V}\chemV\chemV' - 16 {\cal Z} \chemU^\prime \chemV 
+ \tfrac13(1+\chemM)\chemV''' - \tfrac23\chemM^\prime\chemV'' + \chemM''\chemV^\prime \\
\Phi_{uu\varphi}^{(0)} &= {\cal N}\big(-\tfrac13(1+\chemM)\chemV'' + \chemM^\prime\chemV^\prime - 2\chemM''\chemV\big) + 8{\cal Z}\big(\tfrac23\chemV\chemV'' - \tfrac12\chemV^{\prime\,2}\big) \,.
\end{align}
\end{subequations}

\section{Flat space cosmologies with spin-3 chemical potential}\label{se:4.1}

The general result for spin-2 and spin-3 fields, \eqref{eq:cp20}-\eqref{eq:cp28} together with the formulas from appendix \ref{app:B}, is fairly lengthy. It is therefore useful to consider a simple non-trivial class of configurations for applications. In this appendix we achieve this by studying zero mode solutions with most (but not all) chemical potentials switched off. This analysis provides flat space cosmology solutions with spin-3 hair, which can be considered as the flat space analogue of BTZ black holes with spin-3 hair \cite{Ammon:2012wc,Perez:2014pya}.

We consider now zero mode solutions, ${\cal M}^\prime={\cal N}^\prime={\cal V}^\prime={\cal Z}^\prime=0$, with vanishing spin-2 chemical potentials, $\chemM=\chemL=0$, and constant spin-3 chemical potentials, $\chemV^\prime=\chemU^\prime=0$. If we have $\chemU\neq 0$ then $g_{uu}$ acquires a contribution quadratic in the radial coordinate $r$, see \eqref{eq:cp20aa}. Since we want to consider solutions that in the spin-2 sector look like flat space cosmologies \cite{Cornalba:2002fi,Cornalba:2003kd} we must not have such a contribution. Therefore, we switch off the even spin-3 chemical potential as well, $\chemU=0$. In this case entropy \eqref{eq:cp124} simplifies to the spin-2 result \eqref{eq:S2a}.

The metric \eqref{eq:cp20} simplifies to
\eq{
g_{\mu\nu}\,\extd x^\mu\extd x^\nu = \big({\cal M} + 24{\cal V}\chemV +\tfrac43{\cal M}^2\chemV^2\big)\extd u^2
-2\extd r\extd u + \big({\cal L}+8{\cal U}\chemV\big)\,2\extd u\extd\varphi +r^2\extd\varphi^2
}{eq:cp30}
and the spin-3 field \eqref{eq:cp21} simplifies to
\begin{multline}
\Phi_{\mu\nu\lambda} \extd x^\mu \extd x^\nu \extd x^\lambda = \big(2{\cal V} + 64 {\cal V}^2  \chemV^3 + 8 {\cal M} {\cal V} \chemV^2  + \tfrac23 {\cal M}^2 \chemV - \tfrac{8}{27}{\cal M}^3 \chemV^3\big) \extd u^3 \\
-\big(16{\cal V}\chemV^2 + \tfrac43{\cal M}\chemV\big)\extd r \extd u^2+\big(4{\cal U} - \tfrac{16}{3}({\cal M}{\cal U}-3{\cal V}{\cal L})\chemV^2  + \tfrac43{\cal M}{\cal L}\chemV  \big) \extd u^2 \extd\varphi \\
+ \chemV\,\extd r^2\extd u -2{\cal L}\chemV\,\extd r\extd u \extd\varphi 
+ \tfrac13\,\big(-r^2{\cal M}\chemV+3{\cal L}^2\chemV\big)\,\extd u\extd\varphi^2 \,.
\label{eq:cp31}
\end{multline} 

The metric thus receives a contribution from the spin-3 charges ${\cal V}$ and ${\cal U}$, by contrast to what happens in the absence of a spin-3 chemical potential \cite{Afshar:2013vka,Gonzalez:2013oaa}. Switching on a spin-3 chemical potential therefore leads to deformed geometries, some of which can be interpreted as flat space cosmologies with spin-3 hair. 

More specifically, flat space cosmologies with mass parameter $m$ and angular momentum parameter $j$, 
\eq{
\extd s^2 = m\extd u^2 -2\extd r \extd u + 2j\extd u\extd\varphi + r^2 \extd\varphi^2
}{eq:cp48}
are obtained for the choices
\eq{
{\cal V} = \frac{3 (m-{\cal M}) - 4{\cal M}^2 \chemV^2}{72 \chemV} \qquad {\cal U} = \frac{j-{\cal L}}{8 \chemV}\,.
}{eq:cp36}
Note, however, that these solutions are singular in general, because the holonomy conditions in sections \ref{se:4.2} require ${\cal V}=0$, which uniquely determines the mass parameter $m$ for regular solutions as
\eq{
 m = {\cal M} + \tfrac43{\cal M}^2 \,\chemV^2\,.
}{eq:cp200}
Similarly, the last equation \eqref{eq:cp108} together with \eqref{eq:cp36} determines the angular momentum parameter $j$ for regular solutions as
\eq{
 j = {\cal L} - \tfrac43{\cal M}^2 \,\frac{\chemV^2}{\chemL}\,.
}{eq:cp201}

\section{Chemically odd configurations}\label{se:4.3}

\newcommand{\cp}{\nu}

If we keep only the odd chemical potentials, $\chemM\neq 0\neq \chemV$, and switch off the even ones, $\chemL=0=\chemU$, then the connection \eqref{eq:cp17} simplifies considerably. In particular, the component $a_u$ now only contains odd generators. This feature permits us to consider a simple generalization where $a_\varphi$ is deformed.

Namely, we replace the connection components \eqref{eq:cp17} by
\eq{
a_u = a_u^{(0)}  + a_u^{(\chemM)} + a_u^{(\chemV)}\qquad a_\varphi = a_\varphi^{(0)} + a_\varphi^{(\cp)}
}{eq:cp41}
with the same expressions \eqref{eq:cp18}, \eqref{eq:cp19} as before and with the additional term
\eq{
a_\varphi^{(\cp)} = \sum_{n=-1}^1 m_n(\varphi) M_n+\sum_{n=-2}^2 v_n(\varphi) V_n\,.
}{eq:cp40}
The additional term $a_\varphi^{(\cp)}$ commutes with the group element $b$ as defined in \eqref{eq:cp5} and with all contributions to $a_u$, since all commutators involve exclusively two odd generators. Moreover, the expression $\extd a^{(\cp)}$ vanishes since $a^{(\cp)}$ has only a $\varphi$-component and all the functions therein depend on $\varphi$ only. Therefore, the additional term \eqref{eq:cp40} does not contribute to gauge curvature and the full connection \eqref{eq:cp4}-\eqref{eq:cp6} with \eqref{eq:cp41}, \eqref{eq:cp7}, \eqref{eq:cp18}, \eqref{eq:cp19} and \eqref{eq:cp40} solves the Chern--Simons field equations \eqref{eq:cp8}.

The asymptotic behavior of the metric and spin-3 field remain essentially unchanged, so that it may be tempting to consider these generalized flat space solutions of the equations of motion as legitimate field configurations. However, as we now show the canonical charges are in general not well-defined, unless there are some further restrictions on the functions $m_n$ and $v_n$ in \eqref{eq:cp40}. 

The boundary condition preserving gauge transformations do acquire additional terms, $\eps^{(0)}+\De\eps^{(0)}$, as compared to \eqref{eq:cp24}. 
\begin{align}
\De\eps^{(0)} &= \big(v_1{\cal M}\chi+\tfrac13v_2(5{\cal M}\chi^\prime+2{\cal M}^\prime\chi)\big)\,M_0 + \De\eps_{M_{-1}}\,M_{-1} \nonumber \\
&\quad + \big(\tfrac12m_1{\cal M}\chi - 8v_2{\cal V}\chi + \tfrac12v_2{\cal M}\epsilon \big)\,V_0 + \De\eps_{V_{-1}}\,V_{-1} + \De\eps_{V_{-2}}\, V_{-2}
 \label{eq:56}
\end{align}
with
\begin{align}
\De\eps_{M_{-1}} &= \tfrac14m_1({\cal M}\epsilon+16{\cal V}\chi)  
+ (\tfrac12v_2{\cal M}^2 - \tfrac12v_2^\prime{\cal M} - \tfrac13v_2^\prime{\cal M}^\prime  - \tfrac23v_2{\cal M}'') \chi \nonumber \\
&\quad - (\tfrac56v_2^\prime{\cal M} + \tfrac73v_2{\cal M}^\prime)\chi^\prime - \tfrac{13}{6}v_2{\cal M}\chi'' + 4v_2{\cal V}\epsilon - \tfrac23v_1{\cal M}^\prime\chi  - \tfrac{11}{12}v_1{\cal M}\chi^\prime  \displaybreak[1] \\
\De\eps_{V_{-1}} &= - \tfrac{7}{12}m_1{\cal M}\chi^\prime  - 2m_1{\cal M}^\prime\chi - \tfrac16m_1^\prime{\cal M}\chi + 4v_1{\cal V}\chi + \tfrac14v_1{\cal M}\epsilon  \nonumber \\
&\quad - 32(v_2{\cal V})^\prime)\chi  - 16v_2{\cal V}^\prime\chi   -\tfrac16 v_2{\cal M}\epsilon^\prime  - v_2{\cal M}^\prime\epsilon  - \tfrac16v_2^\prime{\cal M}\epsilon  \displaybreak[1] \\
\De\eps_{V_{-2}} &= \tfrac{1}{48}\big(-2m_0 {\cal M}^\prime \chi - 5m_0 {\cal M} \chi^\prime - 6m_1 {\cal M}^2 \chi +  6m_1 {\cal M}''\chi + 15m_1 {\cal M} \chi'' \nonumber \\
&\quad  + 18m_1 {\cal M}^\prime \chi^\prime - 
 24m_1 {\cal V} \epsilon + 6v_0 (16 {\cal V} \chi + {\cal M} \epsilon) - 48v_1 {\cal V}^\prime \chi - 48v_1 {\cal V} \chi^\prime - 3v_1 {\cal M}^\prime \epsilon \nonumber \\
&\quad - 3v_1 {\cal M} \epsilon^\prime - 12m_{-1} {\cal M} \chi + v_2 (32 {\cal V}'' \chi + 32 {\cal V} \chi'' + 64 {\cal V}^\prime \chi^\prime - 
   3 {\cal M}^2 \epsilon + 2 {\cal M}'' \epsilon + 4 {\cal M}^\prime \epsilon^\prime \nonumber \\
&\quad +  2 {\cal M} (-24 {\cal V} \chi + \epsilon'')) + 2m_1'' {\cal M} \chi + 6 m_1^\prime {\cal M}^\prime \chi + 9m_1^\prime {\cal M} \chi^\prime - 48v_1^\prime {\cal V} \chi - 3 v_1^\prime {\cal M} \epsilon \nonumber \\
&\quad + 32v_2'' {\cal V} \chi +  2 v_2'' {\cal M} \epsilon + 64v_2^\prime {\cal V}^\prime \chi + 64v_2^\prime {\cal V} \chi^\prime + 4v_2^\prime {\cal M}^\prime \epsilon + 4v_2^\prime {\cal M} \epsilon^\prime \big)
 \label{eq:57}
\end{align}
Insertion of $\eps^{(0)}+\De\eps^{(0)}$ into the general result for the canonical currents \eqref{eq:cp22} then yields
\begin{multline}
\delta \widehat Q[\varepsilon] = \delta Q[\varepsilon] + \frac{k}{4\pi}\,\oint\extd\varphi\,\big(2{\cal M}\de m_1 \epsilon + 16{\cal V}\de v_2 \epsilon - 8\de m_{-1}\epsilon - 4\de m_0\epsilon^\prime - 4\de m_1 \epsilon'' \\
 - \tfrac83{\cal M}\de v_0 \chi - \tfrac{10}{3}{\cal M}\de v_1 \chi^\prime - \tfrac43 {\cal M}^\prime\de v_1\chi  - \tfrac{16}{3}{\cal M}\de v_2\chi''  - \tfrac{14}{3}{\cal M}^\prime\de v_2 \chi^\prime - \tfrac43 {\cal M}''\de v_2 \chi + 2{\cal M}^2\de v_2\chi \\
+ 32{\cal V}\de m_1\chi + 32\de v_{-2}\chi + 8\de v_{-1}\chi^\prime + \tfrac83\de v_0 \chi'' + \tfrac43 \de v_1 \chi''' + \tfrac43\de v_2 \chi'''' \big)
\label{eq:cp39}
\end{multline}
where $\delta Q[\varepsilon]$ is the previous contribution \eqref{eq:cp25}. 
The first two new terms, the whole second line and the first term in the last line are not integrable in general, which means that the canonical charges are not well-defined. 

The simplest way to obtain integrable canonical charges is to demand
\eq{
\delta m_1 = \delta v_2 = \delta v_1 = \delta v_0 = 0\,.
}{eq:cp38}
With the conditions \eqref{eq:cp38} the canonical charges then read
\eq{
\widehat Q[\varepsilon] = Q[\varepsilon] + \frac{k}{\pi}\,\oint\extd\varphi\,\big(-2\de m_{-1}\epsilon - \de m_0\epsilon^\prime  + 8\de v_{-2}\chi + 2\de v_{-1}\chi^\prime  \big)\,.
}{eq:cp37}
Note that the canonical charges change if we allow any of the quantities $m_n$ or $v_n$ to be state-dependent functions. The quantities $m_0$ and $m_{-1}$ can be absorbed by redefinitions of the state-dependent functions and suitable diffeomorphisms. This can be seen as follows. 

For simplicity, let us assume ${\cal V}={\cal Z}=\mu_V=0$.
Switching on $m_0(\varphi)$ and $m_{-1}(\varphi)$ in the deformation of $a_\varphi$ \eqref{eq:cp40} leads to the line-element
\begin{multline}
g_{\mu\nu}\,\extd x^\mu\extd x^\nu = \big({\cal M}(1+\chemM)^2 - 2(1+\chemM)\chemM'' + (\chemM^\prime)^2 \big)\extd u^2 - \big(1+\chemM\big)\,2\extd r\extd u \\
+ \big(({\cal N}-2m_{-1})(1+\chemM) -  (r+m_0)\chemM^\prime \big)\,2\extd u\extd\varphi + (r+m_0)^2\extd\varphi^2 \,.
\label{eq:cp51}
\end{multline}
The associated canonical charges are given by
\eq{
\widehat Q[\epsilon,\tau] = \frac{k}{\pi}\,\oint\extd\varphi\,\big(({\cal L}-2 m_{-1} + m_0^\prime)\epsilon + {\cal M} \tau \big)\,.
}{eq:cp52}
The coordinate transformation $r+m_0\to r$ together with the redefinition of the function ${\cal L}+m_0^\prime \to {\cal L}$ allows to eliminate the function $m_0$ from the line-element \eqref{eq:cp51} and the canonical charges \eqref{eq:cp52}. A redefinition of the function ${\cal L}-2m_{-1} \to {\cal L}$ eliminates the function $m_{-1}$ from the line-element \eqref{eq:cp51} and the canonical charges \eqref{eq:cp52}. Therefore, the functions $m_0$ and $m_{-1}$ play no physical role.
We expect that essentially the same is true for the quantities $v_{-1}$ and $v_{-2}$, replacing diffeomorphisms by spin-3 gauge transformations.

Note, however, that there are more complicated ways to obtain integrable charges than demanding \eqref{eq:cp38}. We do not study this issue exhaustively here, but just provide one non-trivial example. Choosing
\eq{
m_1 = \nu(\varphi)\,{\cal M}\qquad v_0 = -12\nu(\varphi)\,{\cal V} + f({\cal M})\qquad \delta v_{\pm 2} = \delta v_{\pm 1}  = \delta m_0 = \delta m_{-1} = 0
}{eq:cp58}
we obtain integrable canonical charges
\eq{
\widehat Q[\varepsilon] = Q[\varepsilon] + \frac{k}{\pi}\,\oint\extd\varphi\,\nu(\varphi)\big(\tfrac14\,{\cal M}^2\,\epsilon 
 - {\cal M}\epsilon'' + 8{\cal M}{\cal V}\, \chi - 8{\cal V} \chi'' \big) +  Q_f[\varepsilon] 
}{eq:cp59}
with
\eq{
Q_f[\varepsilon] =  \frac{2k}{3\pi}\,\oint\extd\varphi\,\bigg(f({\cal M})\chi'' - \int^{\cal M}\!\!\!\!\!\!\extd m\,m\,\frac{\extd f(m)}{\extd m}\, \chi \bigg)\,.
}{eq:cp69}

\end{appendix}

\linespread{1.28}

\addcontentsline{toc}{section}{References} 

\begin{thebibliography}{100}

\bibitem{Kapusta}
J.~Kapusta, {\em Finite-Temperature Field Theory: Principles and Applications}.
\newblock Cambridge Monographs on Mathematical Physics, 2006.

\bibitem{Klebanov:2002ja}
I.~Klebanov and A.~Polyakov, {\it {AdS dual of the critical O(N) vector
  model}},  {\em Phys.Lett.} {\bf B550} (2002) 213--219
  [\href{http://arXiv.org/abs/hep-th/0210114}{{\tt hep-th/0210114}}].

\bibitem{Mikhailov:2002bp}
A.~Mikhailov, {\it {Notes on higher spin symmetries}},
  \href{http://arXiv.org/abs/hep-th/0201019}{{\tt hep-th/0201019}}.

\bibitem{Sezgin:2002rt}
E.~Sezgin and P.~Sundell, {\it {Massless higher spins and holography}},  {\em
  Nucl.Phys.} {\bf B644} (2002) 303--370
  [\href{http://arXiv.org/abs/hep-th/0205131}{{\tt hep-th/0205131}}].

\bibitem{Fradkin:1987ks}
E.~Fradkin and M.~A. Vasiliev, {\it {On the Gravitational Interaction of
  Massless Higher Spin Fields}},  {\em Phys.Lett.} {\bf B189} (1987) 89--95.

\bibitem{Fradkin:1986qy}
E.~Fradkin and M.~A. Vasiliev, {\it {Cubic Interaction in Extended Theories of
  Massless Higher Spin Fields}},  {\em Nucl.Phys.} {\bf B291} (1987) 141.

\bibitem{Vasiliev:1990en}
M.~A. Vasiliev, {\it {Consistent equation for interacting gauge fields of all
  spins in (3+1)-dimensions}},  {\em Phys.Lett.} {\bf B243} (1990) 378--382.

\bibitem{Sagnotti:2010at}
A.~Sagnotti and M.~Taronna, {\it {String Lessons for Higher-Spin
  Interactions}},  {\em Nucl.Phys.} {\bf B842} (2011) 299--361
  [\href{http://arXiv.org/abs/1006.5242}{{\tt 1006.5242}}].

\bibitem{Vasiliev:2012vf}
M.~A. Vasiliev, {\it {Holography, Unfolding and Higher-Spin Theory}},  {\em
  J.Phys.} {\bf A46} (2013) 214013 [\href{http://arXiv.org/abs/1203.5554}{{\tt
  1203.5554}}].

\bibitem{Didenko:2014dwa}
V.~Didenko and E.~Skvortsov, {\it {Elements of Vasiliev theory}},
  \href{http://arXiv.org/abs/1401.2975}{{\tt 1401.2975}}.

\bibitem{Giombi:2009wh}
S.~Giombi and X.~Yin, {\it {Higher Spin Gauge Theory and Holography: The
  Three-Point Functions}},  {\em JHEP} {\bf 1009} (2010) 115
  [\href{http://arXiv.org/abs/0912.3462}{{\tt 0912.3462}}].

\bibitem{Giombi:2010vg}
S.~Giombi and X.~Yin, {\it {Higher Spins in AdS and Twistorial Holography}},
  {\em JHEP} {\bf 1104} (2011) 086 [\href{http://arXiv.org/abs/1004.3736}{{\tt
  1004.3736}}].

\bibitem{Koch:2010cy}
R.~d.~M. Koch, A.~Jevicki, K.~Jin and J.~P. Rodrigues, {\it {$AdS_4/CFT_3$
  Construction from Collective Fields}},  {\em Phys.Rev.} {\bf D83} (2011)
  025006 [\href{http://arXiv.org/abs/1008.0633}{{\tt 1008.0633}}].

\bibitem{Giombi:2011ya}
S.~Giombi and X.~Yin, {\it {On Higher Spin Gauge Theory and the Critical O(N)
  Model}},  {\em Phys.Rev.} {\bf D85} (2012) 086005
  [\href{http://arXiv.org/abs/1105.4011}{{\tt 1105.4011}}].

\bibitem{Douglas:2010rc}
M.~R. Douglas, L.~Mazzucato and S.~S. Razamat, {\it {Holographic dual of free
  field theory}},  {\em Phys.Rev.} {\bf D83} (2011) 071701
  [\href{http://arXiv.org/abs/1011.4926}{{\tt 1011.4926}}].

\bibitem{Maldacena:2011jn}
J.~Maldacena and A.~Zhiboedov, {\it {Constraining Conformal Field Theories with
  A Higher Spin Symmetry}},  {\em J.Phys.} {\bf A46} (2013) 214011
  [\href{http://arXiv.org/abs/1112.1016}{{\tt 1112.1016}}].

\bibitem{Maldacena:2012sf}
J.~Maldacena and A.~Zhiboedov, {\it {Constraining conformal field theories with
  a slightly broken higher spin symmetry}},  {\em Class.Quant.Grav.} {\bf 30}
  (2013) 104003 [\href{http://arXiv.org/abs/1204.3882}{{\tt 1204.3882}}].

\bibitem{Maldacena:1997re}
J.~M. Maldacena, {\it {The large N limit of superconformal field theories and
  supergravity}},  {\em Adv. Theor. Math. Phys.} {\bf 2} (1998) 231--252
  [\href{http://arXiv.org/abs/hep-th/9711200}{{\tt hep-th/9711200}}].

\bibitem{Gubser:1998bc}
S.~S. Gubser, I.~R. Klebanov and A.~M. Polyakov, {\it Gauge theory correlators
  from non-critical string theory},  {\em Phys. Lett.} {\bf B428} (1998)
  105--114 [\href{http://arXiv.org/abs/hep-th/9802109}{{\tt hep-th/9802109}}].

\bibitem{Witten:1998qj}
E.~Witten, {\it {Anti-de Sitter space and holography}},  {\em Adv. Theor. Math.
  Phys.} {\bf 2} (1998) 253--291
  [\href{http://arXiv.org/abs/hep-th/9802150}{{\tt hep-th/9802150}}].

\bibitem{Achucarro:1986vz}
A.~Achucarro and P.~K. Townsend, {\it A {C}hern-{S}imons action for
  three-dimensional {A}nti-de {S}itter supergravity theories},  {\em Phys.
  Lett.} {\bf B180} (1986) 89.

\bibitem{Witten:1988hc}
E.~Witten, {\it (2+1)-dimensional gravity as an exactly soluble system},  {\em
  Nucl. Phys.} {\bf B311} (1988) 46.

\bibitem{Blencowe:1988gj}
M.~Blencowe, {\it A consistent interacting massless higher spin field theory in
  d = (2+1)},  {\em Class.Quant.Grav.} {\bf 6} (1989) 443.

\bibitem{Gutperle:2011kf}
M.~Gutperle and P.~Kraus, {\it {Higher Spin Black Holes}},  {\em JHEP} {\bf
  1105} (2011) 022 [\href{http://arXiv.org/abs/1103.4304}{{\tt 1103.4304}}].

\bibitem{Ammon:2011nk}
M.~Ammon, M.~Gutperle, P.~Kraus and E.~Perlmutter, {\it {Spacetime Geometry in
  Higher Spin Gravity}},  {\em JHEP} {\bf 1110} (2011) 053
  [\href{http://arXiv.org/abs/1106.4788}{{\tt 1106.4788}}].

\bibitem{Compere:2013gja}
G.~Comp\`ere and W.~Song, {\it {$\mathcal{W}$ symmetry and integrability of
  higher spin black holes}},  {\em JHEP} {\bf 1309} (2013) 144
  [\href{http://arXiv.org/abs/1306.0014}{{\tt 1306.0014}}].

\bibitem{Compere:2013nba}
G.~Comp{\`e}re, J.~I. Jottar and W.~Song, {\it {Observables and Microscopic
  Entropy of Higher Spin Black Holes}},  {\em JHEP} {\bf 1311} (2013) 054
  [\href{http://arXiv.org/abs/1308.2175}{{\tt 1308.2175}}].

\bibitem{Henneaux:2013dra}
M.~Henneaux, A.~Perez, D.~Tempo and R.~Troncoso, {\it {Chemical potentials in
  three-dimensional higher spin anti-de Sitter gravity}},  {\em JHEP} {\bf
  1312} (2013) 048 [\href{http://arXiv.org/abs/1309.4362}{{\tt 1309.4362}}].

\bibitem{Bunster:2014mua}
C.~Bunster, M.~Henneaux, A.~Perez, D.~Tempo and R.~Troncoso, {\it {Generalized
  Black Holes in Three-dimensional Spacetime}},  {\em JHEP} {\bf 1405} (2014)
  031 [\href{http://arXiv.org/abs/1404.3305}{{\tt 1404.3305}}].

\bibitem{Gaberdiel:2010pz}
M.~R. Gaberdiel and R.~Gopakumar, {\it {An AdS$_3$ Dual for Minimal Model
  CFTs}},  {\em Phys.Rev.} {\bf D83} (2011) 066007
  [\href{http://arXiv.org/abs/1011.2986}{{\tt 1011.2986}}].

\bibitem{Ammon:2012wc}
M.~Ammon, M.~Gutperle, P.~Kraus and E.~Perlmutter, {\it {Black holes in three
  dimensional higher spin gravity: A review}},  {\em J.Phys.} {\bf A46} (2013)
  214001 [\href{http://arXiv.org/abs/1208.5182}{{\tt 1208.5182}}].

\bibitem{Gaberdiel:2012uj}
M.~R. Gaberdiel and R.~Gopakumar, {\it {Minimal Model Holography}},  {\em
  J.Phys.} {\bf A46} (2013) 214002 [\href{http://arXiv.org/abs/1207.6697}{{\tt
  1207.6697}}].

\bibitem{Perez:2014pya}
A.~Perez, D.~Tempo and R.~Troncoso, {\it {Brief review on higher spin black
  holes}},  \href{http://arXiv.org/abs/1402.1465}{{\tt 1402.1465}}.

\bibitem{Afshar:2014rwa}
H.~Afshar, A.~Bagchi, S.~Detournay, D.~Grumiller, S.~Prohazka and M.~Riegler,
  {\it {Holographic Chern-Simons Theories}},  {\em Lect.Notes Phys.} {\bf 892}
  (2015) 311--329 [\href{http://arXiv.org/abs/1404.1919}{{\tt 1404.1919}}].

\bibitem{Krishnan:2013zya}
C.~Krishnan, A.~Raju, S.~Roy and S.~Thakur, {\it {Higher Spin Cosmology}},
  {\em Phys.Rev.} {\bf D89} (2014) 045007
  [\href{http://arXiv.org/abs/1308.6741}{{\tt 1308.6741}}].

\bibitem{Gary:2012ms}
M.~Gary, D.~Grumiller and R.~Rashkov, {\it {Towards non-AdS holography in
  3-dimensional higher spin gravity}},  {\em JHEP} {\bf 1203} (2012) 022
  [\href{http://arXiv.org/abs/1201.0013}{{\tt 1201.0013}}].

\bibitem{Afshar:2012nk}
H.~Afshar, M.~Gary, D.~Grumiller, R.~Rashkov and M.~Riegler, {\it {Non-AdS
  holography in 3-dimensional higher spin gravity - General recipe and
  example}},  {\em JHEP} {\bf 1211} (2012) 099
  [\href{http://arXiv.org/abs/1209.2860}{{\tt 1209.2860}}].

\bibitem{Afshar:2012hc}
H.~Afshar, M.~Gary, D.~Grumiller, R.~Rashkov and M.~Riegler, {\it
  {Semi-classical unitarity in 3-dimensional higher-spin gravity for
  non-principal embeddings}},  {\em Class. Quant. Grav.} {\bf 30} (2012) 104004
  [\href{http://arXiv.org/abs/1211.4454}{{\tt 1211.4454}}].

\bibitem{Gutperle:2013oxa}
M.~Gutperle, E.~Hijano and J.~Samani, {\it {Lifshitz black holes in higher spin
  gravity}},  {\em JHEP} {\bf 1404} (2014) 020
  [\href{http://arXiv.org/abs/1310.0837}{{\tt 1310.0837}}].

\bibitem{Gary:2014mca}
M.~Gary, D.~Grumiller, S.~Prohazka and S.-J. Rey, {\it {Lifshitz Holography
  with Isotropic Scale Invariance}},  {\em JHEP} {\bf 1408} (2014) 001
  [\href{http://arXiv.org/abs/1406.1468}{{\tt 1406.1468}}].

\bibitem{Afshar:2013vka}
H.~Afshar, A.~Bagchi, R.~Fareghbal, D.~Grumiller and J.~Rosseel, {\it {Higher
  spin theory in 3-dimensional flat space}},  {\em Phys.Rev.Lett.} {\bf 111}
  (2013) 121603 [\href{http://arXiv.org/abs/1307.4768}{{\tt 1307.4768}}].

\bibitem{Gonzalez:2013oaa}
H.~A. Gonzalez, J.~Matulich, M.~Pino and R.~Troncoso, {\it {Asymptotically flat
  spacetimes in three-dimensional higher spin gravity}},  {\em JHEP} {\bf 1309}
  (2013) 016 [\href{http://arXiv.org/abs/1307.5651}{{\tt 1307.5651}}].

\bibitem{'tHooft:1993gx}
G.~'t~Hooft, {\it Dimensional reduction in quantum gravity},  in {\em
  Salamfestschrift}, World Scientific, 1993.
\newblock \href{http://arXiv.org/abs/gr-qc/9310026}{{\tt gr-qc/9310026}}.

\bibitem{Susskind:1995vu}
L.~Susskind, {\it {The World as a hologram}},  {\em J. Math. Phys.} {\bf 36}
  (1995) 6377--6396 [\href{http://arXiv.org/abs/hep-th/9409089}{{\tt
  hep-th/9409089}}].

\bibitem{Coleman:1967ad}
S.~R. Coleman and J.~Mandula, {\it All possible symmetries of the {S} matrix},
  {\em Phys.Rev.} {\bf 159} (1967) 1251--1256.

\bibitem{Aragone:1979hx}
C.~Aragone and S.~Deser, {\it {Consistency Problems of Hypergravity}},  {\em
  Phys.Lett.} {\bf B86} (1979) 161.

\bibitem{Weinberg:1980kq}
S.~Weinberg and E.~Witten, {\it {Limits on Massless Particles}},  {\em
  Phys.Lett.} {\bf B96} (1980) 59.

\bibitem{Bekaert:2010hw}
X.~Bekaert, N.~Boulanger and P.~Sundell, {\it {How higher-spin gravity
  surpasses the spin two barrier: no-go theorems versus yes-go examples}},
  {\em Rev.Mod.Phys.} {\bf 84} (2012) 987--1009
  [\href{http://arXiv.org/abs/1007.0435}{{\tt 1007.0435}}].

\bibitem{Grumiller:2014lna}
D.~Grumiller, M.~Riegler and J.~Rosseel, {\it {Unitarity in three-dimensional
  flat space higher spin theories}},  {\em JHEP} {\bf 1407} (2014) 015
  [\href{http://arXiv.org/abs/1403.5297}{{\tt 1403.5297}}].

\bibitem{Barnich:2006av}
G.~Barnich and G.~Compere, {\it {Classical central extension for asymptotic
  symmetries at null infinity in three spacetime dimensions}},  {\em
  Class.Quant.Grav.} {\bf 24} (2007) F15--F23
  [\href{http://arXiv.org/abs/gr-qc/0610130}{{\tt gr-qc/0610130}}].

\bibitem{Bondi:1962}
H.~Bondi, M.~van~der Burg and A.~Metzner, {\it Gravitational waves in general
  relativity {VII.} {W}aves from axi-symmetric isolated systems},  {\em Proc.
  Roy. Soc. London} {\bf A269} (1962) 21--51.

\bibitem{Sachs:1962}
R.~Sachs, {\it Asymptotic symmetries in gravitational theory},  {\em Phys.
  Rev.} {\bf 128} (1962) 2851--2864.

\bibitem{Bagchi:2009my}
A.~Bagchi and R.~Gopakumar, {\it {Galilean Conformal Algebras and AdS/CFT}},
  {\em JHEP} {\bf 0907} (2009) 037 [\href{http://arXiv.org/abs/0902.1385}{{\tt
  0902.1385}}].

\bibitem{Barnich:2010eb}
G.~Barnich and C.~Troessaert, {\it {Aspects of the BMS/CFT correspondence}},
  {\em JHEP} {\bf 1005} (2010) 062 [\href{http://arXiv.org/abs/1001.1541}{{\tt
  1001.1541}}].

\bibitem{Bagchi:2010zz}
A.~Bagchi, {\it {Correspondence between Asymptotically Flat Spacetimes and
  Nonrelativistic Conformal Field Theories}},  {\em Phys.Rev.Lett.} {\bf 105}
  (2010) 171601.

\bibitem{Bagchi:2012yk}
A.~Bagchi, S.~Detournay and D.~Grumiller, {\it {Flat-Space Chiral Gravity}},
  {\em Phys.Rev.Lett.} {\bf 109} (2012) 151301
  [\href{http://arXiv.org/abs/1208.1658}{{\tt 1208.1658}}].

\bibitem{Bagchi:2012xr}
A.~Bagchi, S.~Detournay, R.~Fareghbal and J.~Simon, {\it {Holography of 3d Flat
  Cosmological Horizons}},  {\em Phys. Rev. Lett.} {\bf 110} (2013) 141302
  [\href{http://arXiv.org/abs/1208.4372}{{\tt 1208.4372}}].

\bibitem{Barnich:2012xq}
G.~Barnich, {\it {Entropy of three-dimensional asymptotically flat cosmological
  solutions}},  {\em JHEP} {\bf 1210} (2012) 095
  [\href{http://arXiv.org/abs/1208.4371}{{\tt 1208.4371}}].

\bibitem{Bagchi:2013lma}
A.~Bagchi, S.~Detournay, D.~Grumiller and J.~Simon, {\it {Cosmic Evolution from
  Phase Transition of Three-Dimensional Flat Space}},  {\em Phys.Rev.Lett.}
  {\bf 111} (2013) 181301 [\href{http://arXiv.org/abs/1305.2919}{{\tt
  1305.2919}}].

\bibitem{Barnich:2012aw}
G.~Barnich, A.~Gomberoff and H.~A. Gonzalez, {\it {The Flat limit of three
  dimensional asymptotically anti-de Sitter spacetimes}},  {\em Phys.Rev.} {\bf
  D86} (2012) 024020 [\href{http://arXiv.org/abs/1204.3288}{{\tt 1204.3288}}].

\bibitem{Bagchi:2012cy}
A.~Bagchi and R.~Fareghbal, {\it {BMS/GCA Redux: Towards Flatspace Holography
  from Non-Relativistic Symmetries}},  {\em JHEP} {\bf 1210} (2012) 092
  [\href{http://arXiv.org/abs/1203.5795}{{\tt 1203.5795}}].

\bibitem{Bagchi:2013bga}
A.~Bagchi, {\it {Tensionless Strings and Galilean Conformal Algebra}},  {\em
  JHEP} {\bf 1305} (2013) 141 [\href{http://arXiv.org/abs/1303.0291}{{\tt
  1303.0291}}].

\bibitem{Costa:2013vza}
R.~N.~C. Costa, {\it {Aspects of the zero $\Lambda$ limit in the AdS/CFT
  correspondence}},  \href{http://arXiv.org/abs/1311.7339}{{\tt 1311.7339}}.

\bibitem{Fareghbal:2013ifa}
R.~Fareghbal and A.~Naseh, {\it {Flat-Space Energy-Momentum Tensor from BMS/GCA
  Correspondence}},  {\em JHEP} {\bf 1403} (2014) 005
  [\href{http://arXiv.org/abs/1312.2109}{{\tt 1312.2109}}].

\bibitem{Krishnan:2013tza}
C.~Krishnan and S.~Roy, {\it {Desingularization of the Milne Universe}},  {\em
  Phys.Lett.} {\bf B734} (2014) 92--95
  [\href{http://arXiv.org/abs/1311.7315}{{\tt 1311.7315}}].

\bibitem{Krishnan:2013wta}
C.~Krishnan, A.~Raju and S.~Roy, {\it {A Grassmann path from $AdS_3$ to flat
  space}},  {\em JHEP} {\bf 1403} (2014) 036
  [\href{http://arXiv.org/abs/1312.2941}{{\tt 1312.2941}}].

\bibitem{Bagchi:2013qva}
A.~Bagchi and R.~Basu, {\it {3D Flat Holography: Entropy and Logarithmic
  Corrections}},  {\em JHEP} {\bf 1403} (2014) 020
  [\href{http://arXiv.org/abs/1312.5748}{{\tt 1312.5748}}].

\bibitem{Detournay:2014fva}
S.~Detournay, D.~Grumiller, F.~Sch{\"o}ller and J.~Simon, {\it {Variational
  principle and 1-point functions in 3-dimensional flat space Einstein
  gravity}},  {\em Phys.Rev.} {\bf D89} (2014) 084061
  [\href{http://arXiv.org/abs/1402.3687}{{\tt 1402.3687}}].

\bibitem{Barnich:2014kra}
G.~Barnich and B.~Oblak, {\it {Notes on the BMS group in three dimensions: I.
  Induced representations}},  {\em JHEP} {\bf 1406} (2014) 129
  [\href{http://arXiv.org/abs/1403.5803}{{\tt 1403.5803}}].

\bibitem{Barnich:2014cwa}
G.~Barnich, L.~Donnay, J.~Matulich and R.~Troncoso, {\it {Asymptotic symmetries
  and dynamics of three-dimensional flat supergravity}},  {\em JHEP} {\bf 1408}
  (2014) 071 [\href{http://arXiv.org/abs/1407.4275}{{\tt 1407.4275}}].

\bibitem{Riegler:2014bia}
M.~Riegler, {\it {Flat space limit of (Higher-Spin) Cardy formula}},
  \href{http://arXiv.org/abs/1408.6931}{{\tt 1408.6931}}.

\bibitem{Fareghbal:2014qga}
R.~Fareghbal and A.~Naseh, {\it {Aspects of Flat/CCFT Correspondence}},
  \href{http://arXiv.org/abs/1408.6932}{{\tt 1408.6932}}.

\bibitem{Banados:1992wn}
M.~Ba\~nados, C.~Teitelboim and J.~Zanelli, {\it The black hole in
  three-dimensional space-time},  {\em Phys. Rev. Lett.} {\bf 69} (1992)
  1849--1851 [\href{http://arXiv.org/abs/hep-th/9204099}{{\tt
  hep-th/9204099}}].

\bibitem{Banados:1992gq}
M.~Ba\~nados, M.~Henneaux, C.~Teitelboim and J.~Zanelli, {\it Geometry of the
  (2+1) black hole},  {\em Phys. Rev.} {\bf D48} (1993) 1506--1525
  [\href{http://arXiv.org/abs/gr-qc/9302012}{{\tt gr-qc/9302012}}].

\bibitem{Cornalba:2002fi}
L.~Cornalba and M.~S. Costa, {\it {A New cosmological scenario in string
  theory}},  {\em Phys.Rev.} {\bf D66} (2002) 066001
  [\href{http://arXiv.org/abs/hep-th/0203031}{{\tt hep-th/0203031}}].

\bibitem{Cornalba:2003kd}
L.~Cornalba and M.~S. Costa, {\it {Time dependent orbifolds and string
  cosmology}},  {\em Fortsch.Phys.} {\bf 52} (2004) 145--199
  [\href{http://arXiv.org/abs/hep-th/0310099}{{\tt hep-th/0310099}}].

\bibitem{David:2012iu}
J.~R. David, M.~Ferlaino and S.~P. Kumar, {\it {Thermodynamics of higher spin
  black holes in 3D}},  {\em JHEP} {\bf 1211} (2012) 135
  [\href{http://arXiv.org/abs/1210.0284}{{\tt 1210.0284}}].

\bibitem{Chen:2012ba}
B.~Chen, J.~Long and Y.-N. Wang, {\it {Phase Structure of Higher Spin Black
  Hole}},  {\em JHEP} {\bf 1303} (2013) 017
  [\href{http://arXiv.org/abs/1212.6593}{{\tt 1212.6593}}].

\bibitem{Ferlaino:2013vga}
M.~Ferlaino, T.~Hollowood and S.~P. Kumar, {\it {Asymptotic symmetries and
  thermodynamics of higher spin black holes in AdS3}},  {\em Phys.Rev.} {\bf
  D88} (2013) 066010 [\href{http://arXiv.org/abs/1305.2011}{{\tt 1305.2011}}].

\bibitem{Afshar:2013bla}
H.~R. Afshar, {\it {Flat/AdS boundary conditions in three dimensional conformal
  gravity}},  {\em JHEP} {\bf 1310} (2013) 027
  [\href{http://arXiv.org/abs/1307.4855}{{\tt 1307.4855}}].

\bibitem{Banados:1998gg}
M.~Ba\~nados, {\it {Three-dimensional quantum geometry and black holes}},
  \href{http://arXiv.org/abs/hep-th/9901148}{{\tt hep-th/9901148}}.

\bibitem{Carlip:2005zn}
S.~Carlip, {\it {Conformal field theory, (2+1)-dimensional gravity, and the BTZ
  black hole}},  {\em Class. Quant. Grav.} {\bf 22} (2005) R85--R124
  [\href{http://arXiv.org/abs/gr-qc/0503022}{{\tt gr-qc/0503022}}].

\bibitem{Henneaux:2010xg}
M.~Henneaux and S.-J. Rey, {\it {Nonlinear $W_{infinity}$ as Asymptotic
  Symmetry of Three-Dimensional Higher Spin Anti-de Sitter Gravity}},  {\em
  JHEP} {\bf 1012} (2010) 007 [\href{http://arXiv.org/abs/1008.4579}{{\tt
  1008.4579}}].

\bibitem{Campoleoni:2010zq}
A.~Campoleoni, S.~Fredenhagen, S.~Pfenninger and S.~Theisen, {\it {Asymptotic
  symmetries of three-dimensional gravity coupled to higher-spin fields}},
  {\em JHEP} {\bf 1011} (2010) 007 [\href{http://arXiv.org/abs/1008.4744}{{\tt
  1008.4744}}].

\bibitem{Brown:1986nw}
J.~D. Brown and M.~Henneaux, {\it {Central Charges in the Canonical Realization
  of Asymptotic Symmetries: An Example from Three-Dimensional Gravity}},  {\em
  Commun. Math. Phys.} {\bf 104} (1986) 207--226.

\bibitem{Larsen:1997ge}
F.~Larsen, {\it {A String model of black hole microstates}},  {\em Phys.Rev.}
  {\bf D56} (1997) 1005--1008 [\href{http://arXiv.org/abs/hep-th/9702153}{{\tt
  hep-th/9702153}}].

\bibitem{Cvetic:1997uw}
M.~Cvetic and F.~Larsen, {\it {General rotating black holes in string theory:
  Grey body factors and event horizons}},  {\em Phys.Rev.} {\bf D56} (1997)
  4994--5007 [\href{http://arXiv.org/abs/hep-th/9705192}{{\tt
  hep-th/9705192}}].

\bibitem{Curir:1981uc}
A.~Curir, {\it Remarks on a possible relation between gravitational instantons
  and the spin thermodynamics of a {K}err black hole},  {\em Lett.Nuovo Cim.}
  {\bf 31} (1981) 517--520.

\bibitem{Castro:2012av}
A.~Castro and M.~J. Rodriguez, {\it {Universal properties and the first law of
  black hole inner mechanics}},  {\em Phys.Rev.} {\bf D86} (2012) 024008
  [\href{http://arXiv.org/abs/1204.1284}{{\tt 1204.1284}}].

\bibitem{Detournay:2012ug}
S.~Detournay, {\it {Inner Mechanics of 3d Black Holes}},  {\em Phys.Rev.Lett.}
  {\bf 109} (2012) 031101 [\href{http://arXiv.org/abs/1204.6088}{{\tt
  1204.6088}}].

\bibitem{Banados:2012ue}
M.~Banados, R.~Canto and S.~Theisen, {\it {The Action for higher spin black
  holes in three dimensions}},  {\em JHEP} {\bf 1207} (2012) 147
  [\href{http://arXiv.org/abs/1204.5105}{{\tt 1204.5105}}].

\bibitem{Natsuume:2001ba}
M.~Natsuume, {\it {The Singularity problem in string theory}},
  \href{http://arXiv.org/abs/gr-qc/0108059}{{\tt gr-qc/0108059}}.

\bibitem{Castro:2011fm}
A.~Castro, E.~Hijano, A.~Lepage-Jutier and A.~Maloney, {\it {Black Holes and
  Singularity Resolution in Higher Spin Gravity}},  {\em JHEP} {\bf 1201}
  (2012) 031 [\href{http://arXiv.org/abs/1110.4117}{{\tt 1110.4117}}].

\bibitem{Castro:2011iw}
A.~Castro, R.~Gopakumar, M.~Gutperle and J.~Raeymaekers, {\it {Conical Defects
  in Higher Spin Theories}},  {\em JHEP} {\bf 1202} (2012) 096
  [\href{http://arXiv.org/abs/1111.3381}{{\tt 1111.3381}}].

\bibitem{Kiran:2014kca}
K.~S. Kiran, C.~Krishnan, A.~Saurabh and J.~Simon, {\it {Strings vs Spins on
  the Null Orbifold}},  \href{http://arXiv.org/abs/1408.3296}{{\tt 1408.3296}}.

\bibitem{Horowitz:1990ap}
G.~T. Horowitz and A.~R. Steif, {\it {Singular string solutions with
  nonsingular initial data}},  {\em Phys.Lett.} {\bf B258} (1991) 91--96.

\bibitem{FigueroaO'Farrill:2001nx}
J.~M. Figueroa-O'Farrill and J.~Simon, {\it {Generalized supersymmetric
  fluxbranes}},  {\em JHEP} {\bf 0112} (2001) 011
  [\href{http://arXiv.org/abs/hep-th/0110170}{{\tt hep-th/0110170}}].

\bibitem{Liu:2002ft}
H.~Liu, G.~W. Moore and N.~Seiberg, {\it {Strings in a time dependent
  orbifold}},  {\em JHEP} {\bf 0206} (2002) 045
  [\href{http://arXiv.org/abs/hep-th/0204168}{{\tt hep-th/0204168}}].

\bibitem{Simon:2002ma}
J.~Simon, {\it {The Geometry of null rotation identifications}},  {\em JHEP}
  {\bf 0206} (2002) 001 [\href{http://arXiv.org/abs/hep-th/0203201}{{\tt
  hep-th/0203201}}].

\bibitem{Cardy:1986ie}
J.~L. Cardy, {\it Operator content of two-dimensional conformally invariant
  theories},  {\em Nucl. Phys.} {\bf B270} (1986) 186--204.

\bibitem{Strominger:1997eq}
A.~Strominger, {\it Black hole entropy from near-horizon microstates},  {\em
  JHEP} {\bf 02} (1998) 009 [\href{http://arXiv.org/abs/hep-th/9712251}{{\tt
  hep-th/9712251}}].

\bibitem{Kraus:2011ds}
P.~Kraus and E.~Perlmutter, {\it {Partition functions of higher spin black
  holes and their CFT duals}},  {\em JHEP} {\bf 1111} (2011) 061
  [\href{http://arXiv.org/abs/1108.2567}{{\tt 1108.2567}}].

\bibitem{Gaberdiel:2012yb}
M.~R. Gaberdiel, T.~Hartman and K.~Jin, {\it {Higher Spin Black Holes from
  CFT}},  {\em JHEP} {\bf 1204} (2012) 103
  [\href{http://arXiv.org/abs/1203.0015}{{\tt 1203.0015}}].

\bibitem{deBoer:2013gz}
J.~de~Boer and J.~I. Jottar, {\it {Thermodynamics of higher spin black holes in
  $AdS_3$}},  {\em JHEP} {\bf 1401} (2014) 023
  [\href{http://arXiv.org/abs/1302.0816}{{\tt 1302.0816}}].

\bibitem{Beccaria:2013gaa}
M.~Beccaria and G.~Macorini, {\it {Analysis of higher spin black holes with
  spin-4 chemical potential}},  {\em JHEP} {\bf 1407} (2014) 047
  [\href{http://arXiv.org/abs/1312.5599}{{\tt 1312.5599}}].

\bibitem{Bagchi:2014iea}
A.~Bagchi, R.~Basu, D.~Grumiller and M.~Riegler, {\it {Entanglement entropy in
  Galilean conformal field theories and flat holography}},
  \href{http://arXiv.org/abs/1410.4089}{{\tt 1410.4089}}.

\bibitem{Ammon:2013hba}
M.~Ammon, A.~Castro and N.~Iqbal, {\it {Wilson Lines and Entanglement Entropy
  in Higher Spin Gravity}},  {\em JHEP} {\bf 1310} (2013) 110
  [\href{http://arXiv.org/abs/1306.4338}{{\tt 1306.4338}}].

\bibitem{deBoer:2013vca}
J.~de~Boer and J.~I. Jottar, {\it {Entanglement Entropy and Higher Spin
  Holography in AdS$_3$}},  {\em JHEP} {\bf 1404} (2014) 089
  [\href{http://arXiv.org/abs/1306.4347}{{\tt 1306.4347}}].

\bibitem{Datta:2014uxa}
S.~Datta, J.~R. David, M.~Ferlaino and S.~P. Kumar, {\it {A universal
  correction to higher spin entanglement entropy}},  {\em Phys.Rev.} {\bf D90}
  (2014) 041903 [\href{http://arXiv.org/abs/1405.0015}{{\tt 1405.0015}}].

\end{thebibliography}
\providecommand{\href}[2]{#2}\begingroup\raggedright\endgroup

\end{document}